\documentclass{emulateapj}
\usepackage{graphicx}
\usepackage{bm}
\usepackage{amssymb}
\usepackage{amsmath}
\usepackage{cancel}

\usepackage{epstopdf}
\usepackage{natbib}
\usepackage{epsfig}

 \newcommand{\lya}{Ly$\,\alpha$ }

\newcommand{\avg}[1]{\ensuremath{\langle #1 \rangle}}
\newcommand{\bma}{\begin{math}}
\newcommand{\ema}{\end{math}}
\newcommand{\beq}{\begin{equation}}
\newcommand{\eeq}{\end{equation}}
\newcommand{\beqa}{\begin{eqnarray}}
\newcommand{\eeqa}{\end{eqnarray}}
\newcommand{\bc}{\begin{center}}
\newcommand{\ec}{\end{center}} 
\newcommand{\bit}{\begin{itemize}}
\newcommand{\eit}{\end{itemize}}

\font\BFd=cmmib10
\font\BFt=cmmib10
\font\BFs=cmmib10 scaled 700
\font\BFss=cmmib10 scaled 500

\def\bbox#1{%
\relax\ifmmode
\mathchoice
{{\hbox{\BFd #1}}}
{{\hbox{\BFt #1}}}
{{\hbox{\BFs #1}}}
{{\hbox{\BFss #1}}}
\else \mbox{#1} \fi }

\def\k{{\bbox{k}}}
\def\q{{\bbox{q}}}

\def\x{{\bbox{x}}}
\def\thetab{\pmb{\theta}}

\begin{document}

\title{On Removing Interloper Contamination from Intensity Mapping Power Spectrum Measurements}
\author{Adam Lidz\altaffilmark{1} \& Jessie Taylor\altaffilmark{1}}
\altaffiltext{1} {Department of Physics \& Astronomy, University of Pennsylvania, 209 South 33rd Street, Philadelphia, PA 19104, USA}
\email{alidz@sas.upenn.edu}

\begin{abstract}
Line intensity mapping experiments seek to trace large scale structure by measuring the spatial fluctuations in the combined emission, in some convenient spectral line, from 
individually unresolved galaxies. An important systematic concern for these surveys is line confusion from foreground or background galaxies emitting in other lines that happen to
lie at the same observed frequency as the ``target'' emission line of interest. We develop an approach to separate this ``interloper'' emission at the power spectrum level. If one adopts the redshift
of the target emission line in mapping from observed frequency and angle on the sky to co-moving units, the interloper emission is mapped to the wrong co-moving coordinates. Since the mapping
is different in the line of sight and transverse directions, the interloper contribution to the power spectrum becomes anisotropic, especially if the interloper and target emission are at widely
separated redshifts. This distortion is analogous to the Alcock-Paczynski test, but here the warping arises from assuming the wrong redshift rather than an incorrect cosmological model.
We apply this to the case of a hypothetical [CII] emission survey at $z \sim 7$ and find that the distinctive interloper anisotropy can, in principle, be used to separate strong foreground CO
emission fluctuations. In our models, however, a significantly more sensitive instrument than currently planned is required, although there are large uncertainties in forecasting the
high redshift [CII] emission signal. With upcoming surveys, it may nevertheless be useful to apply this approach after
first masking pixels suspected of containing strong interloper contamination. 
\end{abstract}



\section{Introduction} \label{sec:intro}

Intensity Mapping (IM) is an appealing approach for studying the large scale structure of the universe and for 
characterizing the bulk properties of galaxy populations emitting in various spectral lines across cosmic time. IM observations
forego detecting galaxies individually. Instead, one measures the large-scale spatial fluctuations in the collective emission
from all of the luminous sources emitting in some convenient spectral line or lines (see e.g. 
\citealt{Suginohara:1998ti,Chang:2007xk,Righi:2008br,Visbal10,Gong11,Carilli11,Lidz11,Pullen13,Uzgil:2014pga,Breysse:2014uia,Croft:2015nna,Li:2015gqa,Mashian15,Keating:2015qva}). 
This complements traditional galaxy surveys which target individual objects in that IM surveys are sensitive to the collective emission from {\em all} luminous
sources, while traditional observations are limited to detecting only those sources that lie above survey flux limits. 
IM also probes line emission across
a range of large-scale environments, and is sensitive to the emission from galaxies in underdense voids as well as sources in high density peaks. This is often impossible in a traditional survey, where spanning large-scale environmental variations requires capturing an enormous volume at high sensitivity. 

One potentially powerful application is to the Epoch of Reionization (EoR).  Current evidence suggests that the universe is reionized largely by numerous low-luminosity sources (e.g. \citealt{Robertson:2015uda}), and so it is extremely challenging
to detect most of the ionizing sources individually. However, it may nevertheless be possible to study their collective impact using IM.
In addition, by spanning a large field-of-view at coarse angular resolution while retaining redshift information, IM surveys would be well-matched to redshifted 21 cm observations of the EoR. The cross-correlation
of IM measurements with redshifted 21 cm data sets could then be used to confirm the high redshift origin of a putative 21 cm signal from the EoR \citep{Furlanetto:2006pg,Lidz:2008ry,Lidz11}. Only the high
redshift portion of the redshifted 21 cm signal, and not residual foreground emission, should correlate with the IM data (asides for shared foregrounds). Furthermore, the scale-dependence of the cross-correlation between
the two signals provides a powerful probe of the size of the ionized regions that form around groups of galaxies during reionization \citep{Lidz:2008ry,Lidz11,Gong11}.

One systematic concern with IM measurements relates to foreground interloper emission \citep{Visbal10}. This interloper emission arises from sources residing at lower (or possibly higher) redshifts -- and emitting in
different lines -- than targeted by the IM survey, with the interlopers nevertheless contributing to the specific intensity at the observed wavelengths of interest. Explicitly, suppose the
survey targets an emission line with a rest-frame wavelength of $\lambda_{r,t}$ and a target redshift around $z_t$. The observed wavelength of this emission is $\lambda_{\rm obs} =
\lambda_{r,t} (1 + z_t)$. Clearly an interloper source, emitting in a line with rest wavelength $\lambda_{r,i}$, can emit at the same observed wavelength provided its redshift, $z_i$, satisfies
$1+z_i = \lambda_{r,t} (1 + z_t)/\lambda_{r,i}$. 
One approach to avoid bias 
from interloper emission is to  probe two different emission lines from gas at the same redshift. The cross-correlation between the emission at the two corresponding observed wavelengths will, on average, only pick up
contributions from gas at the target redshift (e.g. \citealt{Visbal10}). Although each of the two observed wavelengths will contain interloper emission, the interlopers will be at widely separated redshifts and so uncorrelated. 
It will likely, however, be valuable to have additional handles to discriminate interloper emission. For one, it may not be feasible for the IM surveys to capture multiple
bright lines from the same emitting gas, since this requires high sensitivity over a broad range of wavelengths. Moreover, it is necessary to clean interloper contamination to measure the
{\em auto spectrum} of the emission fluctuations in a line of interest; this quantity contains information that is not available from the cross spectrum between two lines. Another possibility 
is to mask out regions suspected of containing bright interloper emission, but this may require an additional survey to identify which regions to mask (e.g. \citealt{Silva:2014ira}). The second
survey must span the redshift range of all prominent interloper lines, and trace some quantity that is a good proxy for the interloper line emission. Furthermore, redshift information is required for
all of the tracer galaxies. For some applications, it may be necessary to mask a significant fraction of the observed pixels. Finally, the resulting mask will reflect the clustering of the interloper sources; it must
be deconvolved carefully to avoid introducing any bias in the inferred target emission fluctuations.

Here we develop an alternative approach for separating-out interloper contamination at the power spectrum level. Our starting point is to note that the mappings between
observed wavelength/frequency and angle on the sky to co-moving length scales/wavenumbers are redshift dependent. If we assume the {\em target redshift} in converting between the observed frequencies and angles and co-moving
coordinates, the interloper fluctuations will be mapped to the wrong wavenumbers. Since the remapping is different for the line of sight and transverse wavenumbers, the interloper
contribution to the observed power spectrum will have a distinctive anisotropy. This is analogous to the Alcock-Paczynski (AP) effect \citep{Alcock79,Ballinger96}, except in the case of the AP test a warping arises from assuming the wrong cosmology, while here the distortion results from adopting the incorrect redshift. 
We will show that this transfer of power and warping can be used to separate out the interloper contamination. This basic idea is mentioned in previous work by
\citet{Visbal10} and \citet{Gong:2013xda}, but we develop the technique further here and apply it to quantify the prospects for  cleaning interloper lines from future $z \sim 7$ [CII] surveys.
Although we focus on the illustrative example of IM with the [CII] line, our approach should be broadly applicable to IM surveys in other lines such as \lya and CO transitions, and may also be of interest for traditional
surveys detecting line-emitting galaxies. 

The outline of this paper is as follows. In \S \ref{sec:disto}, we describe and quantify the interloper distortion. This is then applied to the example case of a futuristic $z \sim 7$ [CII] emission survey (\S \ref{sec:cii_examp}). \S \ref{sec:forecasts} forecasts the constraints on [CII] and CO emission line properties that may be achieved by this survey. We further consider combining our technique with additional tracers of large-scale structure at the redshifts of prominent foreground interlopers (\S \ref{sec:xcorr_lss}). We also discuss the prospects for cross-correlating with
other emission lines at $z \sim 7$ (\S \ref{sec:xcorr_lines}). We conclude in \S \ref{sec:conclusions}. Throughout we adopt a cosmological model with $\Omega_m = 0.27$, $\Omega_\Lambda = 073$, 
$\Omega_b = 0.046$, $h=0.7$, $\sigma_8(z=0) = 0.8$, and $n_s=1$, broadly consistent with recent Planck measurements
\citep{Ade:2013zuv}.

\section{Interloper Coordinate Mapping Distortions} \label{sec:disto}

In order to illustrate the technique, let us first suppose that our data cube contains only two sources of line emission: our target line of interest at redshift $z_t$, and a single dominant interloper line at redshift
$z_i$. We will soon generalize to the case that several interloper lines contribute.
We denote the observed frequency at the center of the data cube by $\nu_{\rm obs}$ and consider emission offset by a small frequency interval $\Delta \nu_{\rm obs}$ from the
cube center. Further, let $\Delta \thetab$ be the angular separation from the center of the cube; the vector describes the two directions transverse to the line
of sight and we work in the flat sky approximation. In order to convert from the observed $\Delta \nu_{\rm obs}$ and $\Delta \thetab$ to co-moving coordinates, we need
to assume a cosmological model and a redshift for the emission. 

Adopting the target redshift for this mapping will cause the interloper emission to {\em be mapped to the wrong co-moving coordinates}. Let us denote the apparent line of sight coordinate 
for the interloper emission by $\tilde{x}_\parallel$ and the apparent transverse coordinate by $\tilde{\x}_\perp$. Further suppose that the true line of sight and transverse coordinates at the interloper
redshift
are $x_\parallel$ and $\x_\perp$. The apparent coordinates are related to the observable frequency interval and angles by incorrectly assuming the emission is at the target redshift:
\beq
\tilde{x}_\parallel = \frac{c}{H(z_t)} (1 + z_t) \frac{\Delta \nu_{\rm obs}}{\nu_{\rm obs}},
\label{eq:xpar_app}
\eeq
and
\beq
\tilde{\x}_\perp = D_{\rm A, co}(z_t) \Delta \thetab,
\label{eq:xperp_app}
\eeq
where $H(z_t)$ is the Hubble parameter at the target redshift and $D_{\rm A, co}(z_t)$ is the co-moving angular diameter distance to the target redshift. (For a flat universe,
$D_{\rm A,co}(z_t) = \chi(z_t)$ with $\chi(z_t)$  being the co-moving distance to redshift $z_t$.)
The relations between the apparent coordinates,
$\tilde{x}_\parallel$ and $\tilde{\x}_\perp$, and the true coordinates, $x_\parallel$ and $\x_\perp$, are then:
\beq
\tilde{x}_\parallel = \frac{H(z_i)}{H(z_t)} \frac{1+z_t}{1+z_i} x_\parallel,
\label{eq:xpar_warp}
\eeq
and
\beq
\tilde{\x}_\perp = \frac{D_{\rm A, co}(z_t)}{D_{\rm A, co}(z_i)} \x_\perp.
\label{eq:xperp_warp}
\eeq
Since we are ultimately interested in the power spectrum, we also consider the line of sight and transverse components of the co-moving wavenumbers. The relevant factors here are just the inverse of the coordinate mappings:
\beq
\tilde{k}_\parallel = \frac{H(z_t)}{H(z_i)} \frac{1+z_i}{1+z_t} k_\parallel = \alpha_\parallel k_\parallel,
\label{eq:kpar_warp}
\eeq
and
\beq
\tilde{\k}_\perp = \frac{D_{\rm A, co}(z_i)}{D_{\rm A, co}(z_t)} \k_\perp = \alpha_\perp \k_\perp.
\label{eq:kperp_warp}
\eeq
Here we have defined ``distortion'' factors, $\alpha_\parallel$ and $\alpha_\perp$. These describe the remapping
that occurs when the incorrect redshift is used to convert angles and observed frequencies to wavenumbers for the interloper
population.

Turning now to power spectrum, we consider the fluctuations in the specific intensity field, $I_{\rm tot}(\x)$. Note that throughout we will work with this quantity rather than with the 
power spectrum of $\delta_I(\x) = (I_{\rm tot}(\x) - \avg{I_{\rm tot}})/\avg{I_{\rm tot}}$ -- i.e., we don't divide out by $\avg{I_{\rm tot}}$.
The apparent power spectrum of the interloper emission is then:
\beq
\tilde{P}_i(\tilde{k}_\parallel,\tilde{\k}_\perp) = \frac{1}{\alpha_\parallel \alpha_\perp^2} P_i\left(\frac{\tilde{k}_\parallel}{\alpha_\parallel},\frac{\tilde{\k}_\perp}{\alpha_\perp}\right). 
\label{eq:pk_transform}
\eeq
Here $\tilde{P}_i$ is the apparent interloper power spectrum, while $P_i$ is the true interloper power spectrum.
This equation reflects how the power spectrum transforms under a change of coordinates; the $1/(\alpha_\parallel \alpha_\perp^2)$ factor is the ratio of the apparent to actual volume surveyed at the interloper redshift
(see \citealt{Ballinger96} for a related discussion in the context of the AP effect, and \citealt{Visbal10,Gong:2013xda,Pullen:2015yba} for earlier work on interloper contamination).
With this transformation law in hand -- to make our description more compact -- we will generally drop the $(\tilde{k}_\parallel, \tilde{\k}_\perp)$ notation and use $(k_\parallel, \k_\perp)$, nevertheless 
assuming the target redshift to map between wavelength/angle and co-moving units.

The total power spectrum of fluctuations in the specific intensity is then:
\beq
P_{\rm tot}(k_\parallel, \k_\perp) = P_t (k_\parallel, \k_\perp) + \frac{1}{\alpha_\parallel \alpha_\perp^2} P_i \left(\frac{k_\parallel}{\alpha_\parallel},\frac{\k_\perp}{\alpha_\perp}\right).
\label{eq:power_warp}
\eeq
The first term on the right hand side is the underlying ``target'' power spectrum that we seek to determine while the second term arises from the distorted interloper contamination.  
In the case that the target and interloper line redshifts are quite different -- as will often be the case for high redshift intensity mapping observations --  the distortion factors $\alpha_\parallel$ and $\alpha_\perp$ will differ significantly from unity and from each other. Interestingly, provided the target line is at higher redshift than the interloper lines, $\alpha_\parallel$ will be larger than unity, while $\alpha_\perp$ will be
smaller than unity. In other words, the interloper fluctuations that appear at a given $k_\parallel, \k_\perp$ arise from modes that have {\em smaller} line of sight wavenumber and {\em larger} transverse
wavenumber than supposed. Provided $P_i(k_\parallel/\alpha_\parallel, \k_\perp/\alpha_\perp)$ is a decreasing function of $k_\parallel$ and $\k_\perp$, the distortion then enhances the power for line of sight wavemodes relative
to the transverse modes. As we will see, the shifting of power and the anisotropy induced from these coordinate re-mappings may potentially
be used to separate out the target and interloper emission at the power spectrum level. 

\begin{figure}
\begin{center}
\includegraphics[width=9cm]{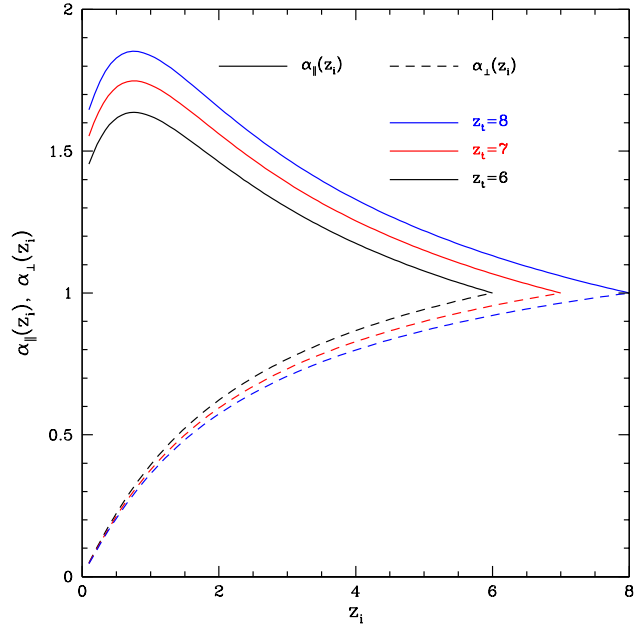}
\caption{Interloper distortion mapping factors, as a function of the interloper redshift, $z_i$. The solid lines show the line-of-sight distortion factor, $\alpha_\parallel(z_i)$, while the
dashed lines show the transverse factor, $\alpha_\perp(z_i)$. The black, red, and blue lines show target redshifts of $z_t=6, 7$, and 8 respectively.}
\label{fig:disto_fac}
\end{center}
\end{figure}

To provide quantatative information, Fig.~\ref{fig:disto_fac} plots the distortion factors as a function of the interloper redshift for a few example target redshifts. Clearly the distortion factors are
quite different from unity and from each other in the case that the target and interloper redshifts are widely separated.  

Naturally, in the more general case that $N$ important interlopers contribute to the power spectrum of fluctuations Eq.~\ref{eq:power_warp} generalizes to:
\begin{align}
P_{\rm tot}(k_\parallel,\k_\perp) =& P_t(k_\parallel,\k_\perp) \nonumber \\
& + \sum_{j=1}^{N} \frac{1}{\alpha_\parallel(z_j) \alpha_\perp^2(z_j)} P_{j} \left(\frac{k_\parallel}{\alpha_\parallel(z_j)},\frac{\k_\perp}{\alpha_\perp(z_j)}\right).
\label{eq:power_warp_tot}
\end{align}
Here the index $j$ denotes the $j$th of the $N$ interloper lines, $z_j$ is the redshift of the $j$th interloper emission line, $P_{j}$ is the specific intensity power spectrum of this emission, and $\alpha_\parallel(z_j)$,
$\alpha_\perp(z_j)$ are the distortion factors which depend on both the redshift of the interloper $z_j$ and the target redshift $z_t$. (We suppress the dependence on the target redshift here to make
the notation less cumbersome.) This equation assumes the interlopers and targets are all widely separated in redshift and so independent of each other (otherwise there would be cross-terms), which should be an extremely good approximation in the case considered below.

\section{Example Application} \label{sec:cii_examp}

Although this technique may have a range of applications, we illustrate it through the interesting example case of a hypothetical survey for [CII] emission at $z_t=7$. Before proceeding further, we very briefly comment on the
physics and phenomenology of the [CII] emission line.
Recall that the ground state configuration
of the five electrons in singly-ionized Carbon is $1s^2 2s^2 2p^1$, and so the ground state has total orbital angular momentum $L=1$ and total spin angular momentum $S=1/2$. The [CII] line is emitted in transitions from the
higher energy fine structure level with total -- orbital plus spin -- angular momentum $J=3/2$ to the lower energy state with $J=1/2$, i.e. it is a $^2 P_{3/2} \rightarrow 2 P_{1/2}$ transition.
The rest-frame wavelength of the transition is $\lambda_r = 157.7 \mu {\rm m}$, the excitation temperature of [CII] is $91$ K, and the energy required to ionize CI to CII is $11.2$ eV.
Since the ionization potential is less than that of neutral hydrogen ($13.6$ eV)
the [CII] emission traces -- in part -- neutral phases of the interstellar medium (ISM), while the low excitation temperature allows emission from warm/cool regions of the ISM. Consequently, [CII] emission may arise from diverse
phases of a galaxy's ISM including photo-dissociation regions at the boundary between molecular clouds and HII regions; from the cold neutral medium; and from HII regions and diffuse ionized gas, provided the local UV radiation
field is insufficiently hard to doubly-ionize carbon (see e.g. the recent review by \citealt{Carilli:2013qm}).

In low redshift galaxies, the [CII] line is a strong cooling line with a luminosity that is $0.1-1\%$ of the total far-infrared luminosity from the galaxy \citep{Stacey91}. Despite the diverse set of conditions
that can give rise to [CII] emission, the line luminosity is fairly well correlated with the star formation rate, at least at low redshift where there are currently good measurements. This is the case even for low-metallicity dwarf galaxies nearby, although the relation shows larger scatter towards
low metallicity \citep{DeLooze:2014dta}. Recent observations have started to detect [CII] emission from Lyman-break selected galaxies and quasar host galaxies at $z \gtrsim 6$, although there
are also a handful of upper limits tentatively suggesting that high redshift galaxies may mostly lie below local [CII] luminosity star-formation rate correlations (e.g. \citealt{Knudsen:2016igf} and references therein).
It is hence unclear how luminous reionization-era
galaxies will be in the [CII] line. Naturally, one of the main goals of IM is to provide a census of the total [CII] emission: while we have much to learn here, this also makes our forecasts uncertain.
In this work we adopt a simplistic approach and assume that local correlations between [CII] luminosity and star-formation rate apply also at high redshift. Likewise, we adopt
local correlations to assess the plausible level of interloper contamination. Future targeted observations of individual galaxies using ALMA will be important
for refining estimates of the target and interloper line luminosities. It may also be instructive to construct models of the interstellar media of high redshift galaxies to try and {\em predict} the correlations
between line luminosity and star formation rate directly (see e.g. \citealt{Munoz:2013tv}).

The central observed wavelength and frequency for our $z_t=7$ [CII] survey are $\lambda_{\rm obs} = 1.26 \times 10^3 \mu {\rm m}$, and $\nu_{\rm obs} = 238$ GHz, respectively. The same observed frequencies will be polluted
with emission from CO molecules at lower redshift undergoing rotational transitions. A CO molecule transitioning between rotational states $J$ and $J-1$ emits a photon of rest-frame frequency $\nu_J = J \times 115$ GHz.  As we will see, several different CO transitions may be significant interlopers for a $z_t=7$ [CII] emission survey. In addition to the CO lines, additional atomic fine structure lines
may also provide non-negligible interloper emission but, as we detail below, these are subdominant to the CO interlopers in our models.

\subsection{Target and Interloper Model Power Spectra}

To proceed, let us first discuss the general form of the model intensity power spectra for both the target and interloper emission.
Incorporating anisotropies from redshift space distortions, our model for the target power spectrum is \citep{Lidz11}:
\begin{align}
P_t(k_\parallel,\k_\perp) =& \avg{I_t}^2 \avg{b_t}^2 \left(1 + \beta_t \mu^2\right)^2 D\left[\mu k \sigma_p(z_i)\right] P_\rho(k,z_t) \nonumber \\ 
& + P_{\rm shot,t}.
\label{eq:ptar_rspace}
\end{align}

Here $\mu = k_\parallel/k$ is the cosine of the angle between the wavevector $\k$ and the line of sight direction, $\avg{I_t}$ is the average specific intensity of the target emission,
and $\avg{b_t}$ is the average luminosity-weighted bias of the emitting galaxies. The factor $\left(1 + \beta_t \mu^2\right)^2$ comes from the Kaiser effect \citep{Kaiser87}, while $D(\mu k \sigma_p)$ quantifies
the small scale reduction of redshift-space power from the finger-of-god effect. The parameter $\beta_t = f_\Omega/\avg{b_t}$ with $f_\Omega = \frac{{\rm dlnD}}{{\rm dlna}}$ denoting the usual logarithmic derivative
of the growth factor, which is well approximated by $f_\Omega \approx \left[\Omega_m(z)\right]^{0.55}$ \citep{Linder:2005in}.  For the finger-of-god suppression, we assume a Lorentzian form:
\beq
D(\mu k \sigma_p) = \frac{1}{1 + \sigma^2_p \mu^2 k^2},
\label{eq:dfog}
\eeq
and approximate the pairwise velocity dispersion by $\sigma_p(z) = \sigma_v(z)/\sqrt{2}$ with $\sigma^2_v(z)$ being the variance of the line-of-sight component of the velocity field according to linear theory. 
In our model, we assume pure linear biasing so that $P_\rho(k,z_t)$ denotes the matter power spectrum according to linear theory. 
Finally, $P_{\rm shot,t}$ is a shot-noise term that arises
because the [CII] emitting galaxies are discrete objects. This term is assumed to be independent of scale. 
Note that we are taking a somewhat simplified model for the redshift-space emission power spectrum: for 
the most part we work on scales much larger than that of individual halos, but we nevertheless include a finger of
god term (owing to virialized motions {\em within} halos). Although this is a bit inconsistent, the measurements we consider are mostly confined
to large scales where the finger-of god suppression and halo profile have negligible impact. In future work, it may be interesting to refine this model (see e.g. \citealt{Cooray02}).

The above equation (Eq. \ref{eq:ptar_rspace}) also highlights another potential benefit of measuring the angular dependence of the power spectrum. Although the first term in this equation depends mostly on the product of
$\avg{I_t}$ and $\avg{b_t}$, there is an additional separate dependence on $\avg{b_t}$ through the parameter $\beta_t$. If the angular dependence of the power spectrum may be measured well enough,
this should help in breaking the otherwise perfect degeneracy between $\avg{I_t}$ and $\avg{b_t}$, and allow one to constrain each of these quantities separately \citep{Lidz11}.

Similarly, the true interloper power spectrum for the $j$th interloper (see Eq. \ref{eq:power_warp_tot}) may be written as a
 function of the true underlying wavenumber components, $k_\parallel$ and $\k_\perp$, as:
\begin{align}
P_{j}(k_\parallel,\k_\perp) =& \avg{I_{j}}^2 \avg{b_{j}}^2 \left(1 + \beta_{j} \mu^2 \right)^2 D\left[\mu k \sigma_p(z_j)\right] P_\rho(k,z_j) \nonumber \\ 
& + P_{\rm shot,j}
\label{eq:pfor_rspace}
\end{align}
The apparent interloper power is
$1/(\alpha_\parallel\alpha^2_\perp) P_j(k_\parallel/\alpha_\parallel, \k_\perp/\alpha_\perp)$, where we momentarily suppress the $j$ indices on the distortion factors. Note that under the coordinate
transformation of Eqs.~\ref{eq:xpar_app}--\ref{eq:kperp_warp} $\mu$ maps to  $\mu \rightarrow (k_\parallel/\alpha_\parallel)/\sqrt{(k_\parallel/\alpha_\parallel)^2 + (k_\perp/\alpha_\perp)^2}$.

Our model for the total power is then specified by the average specific intensity of the target and interloper emission, $\avg{I_t}$ and $\avg{I_{j}}$, the luminosity-weighted average bias factors, $\avg{b_t}$
and $\avg{b_{j}}$, and the shot-noise terms, $P_{\rm shot,t}$ and $P_{\rm shot,j}$. For simplicity, we generally fix $\avg{b_t} = 3$ and $\avg{b_{j}} =2$ (for each interloper $j$) in what follows.
In order to determine plausible values for the average specific intensity and shot-noise terms, we seek guidance 
from empirical correlations between the luminosity in the emission lines of interest and galactic star formation rates.
These correlations can be combined with Schechter function fits to the abundance of galaxies as a function of their star formation rate to estimate the remaining quantities of interest, as 
in \citet{Pullen13}. The Schechter form for the star formation rate function is \citep{Schechter76}:
\beq
\phi({\rm SFR}) d{\rm SFR} = \phi_\star \left(\frac{{\rm SFR}}{{\rm SFR}_\star}\right)^\alpha {\rm exp}\left[-\frac{{\rm SFR}}{{\rm SFR}_\star}\right] \frac{d {\rm SFR}}{{\rm SFR}_\star},
\label{eq:sfr_func}
\eeq
with $\alpha$ denoting the faint-end slope, and ${\rm SFR}_\star$ and $\phi_\star$ giving, respectively, the characteristic star-formation rate and number density.

The average specific intensity in each line can be estimated from the co-moving emissivity in the line according to \citep{Lidz11,Pullen13}:
\beq
\avg{I_L} = \frac{\epsilon_L}{4 \pi \nu_{\rm rest, L}} \frac{c}{H(z)},
\label{eq:i_avg}
\eeq
where $\nu_{\rm rest, L}$ is the restframe emission frequency, $\epsilon_L$ is the co-moving emissivity of the line emission, and the line profile has been approximated as a delta function in frequency. We further approximate the luminosity as a linear function of the star formation rate:
\beq
L = L_0 \frac{{\rm SFR}}{1 M_\odot {\rm yr}^{-1}}.
\label{eq:lum_sfr}
\eeq
Using the Schechter form for the star-formation rate function, it follows that the co-moving emissivity in each line $L$ is \citep{Pullen13}:
\beq
\epsilon_L = \phi_\star L_0 \frac{{\rm SFR}_\star}{1 M_\odot {\rm yr}^{-1}} \Gamma(2 + \alpha).
\label{eq:emiss}
\eeq

\begin{figure}
\begin{center} 
\includegraphics[width=9cm]{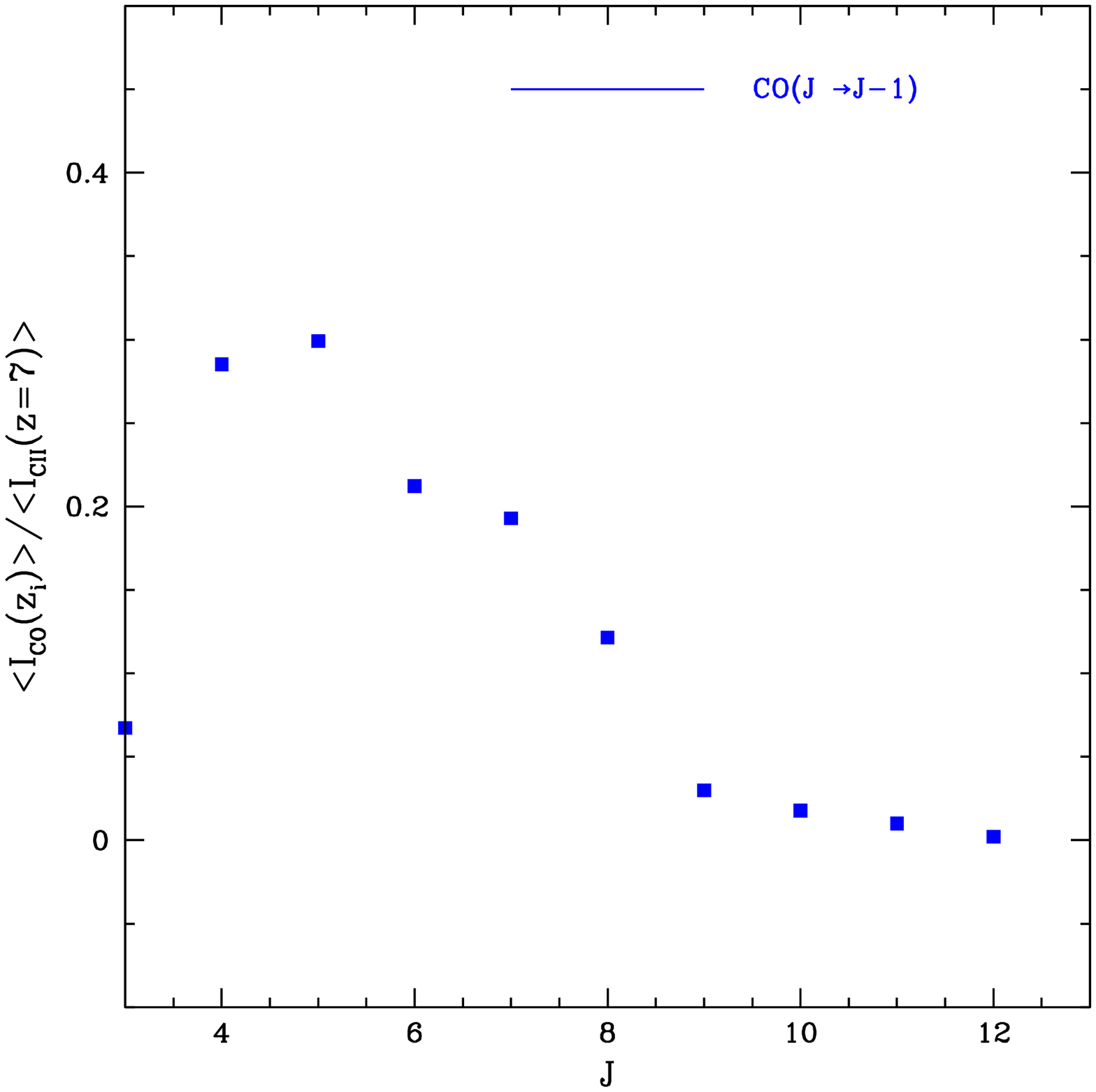}
\includegraphics[width=9cm]{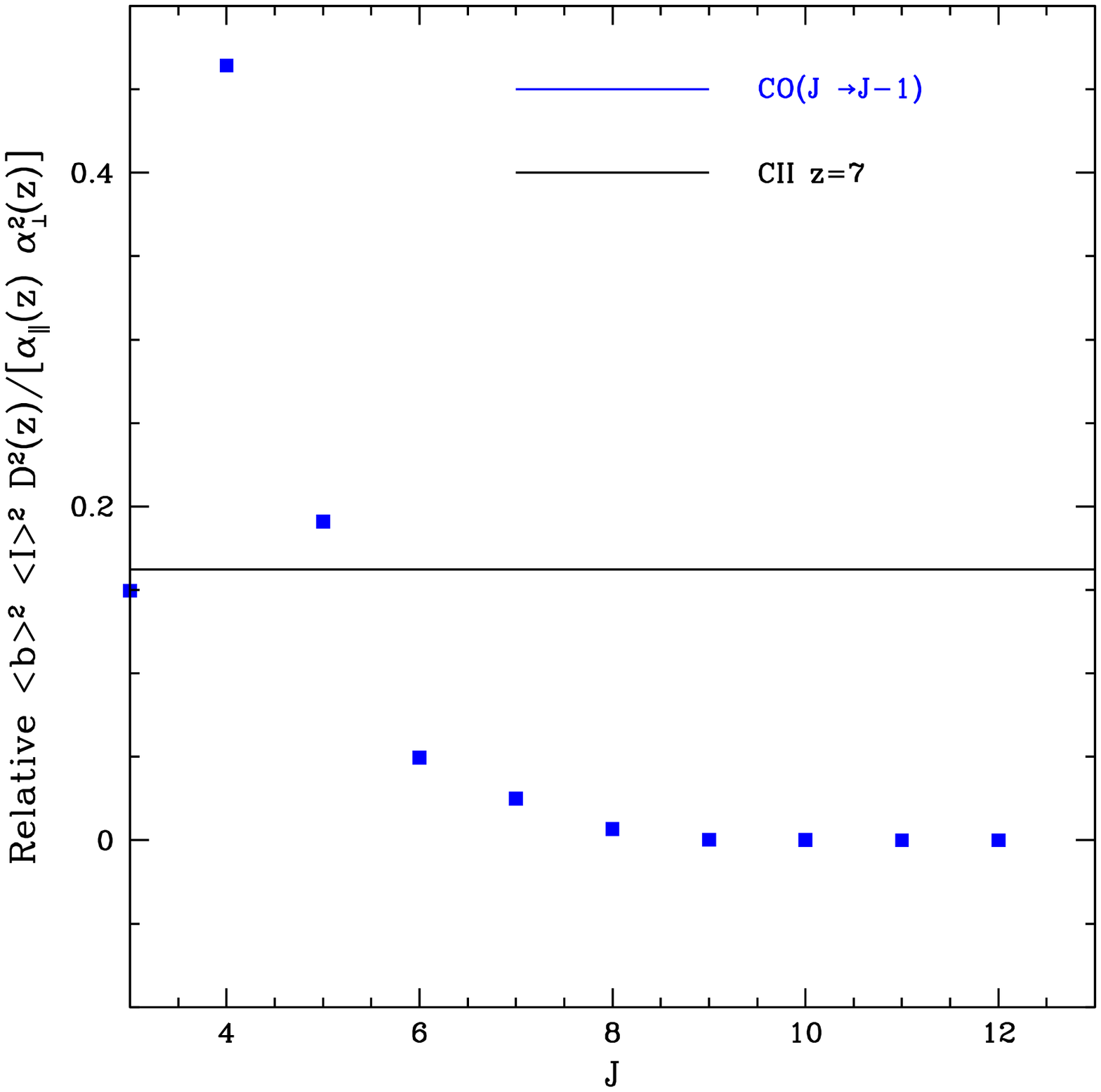}
\caption{Relative importance of interloper lines to the average specific intensity and to the power spectrum of intensity fluctuations. {\em Top}: The estimated intensity of different interloper transitions,
CO$(J \rightarrow J-1)$ as a function of $J$, relative to the model [CII] target emission intensity. {\em Bottom}: The y-axis shows a factor that determines the relative contribution of different interloper
lines to the total power spectrum of intensity fluctuations (see text). The normalization has been set here so that the factor sums (over all lines) to unity. The black horizontal line shows the same
factor for the target line, [CII] at $z=7$.}
\label{fig:interlop_imp}
\end{center}
\end{figure}

We adopt the values of $L^{\rm CII}_0 = 6 \times 10^6 L_\odot$ and the luminosity of the CO transitions given in \citet{Visbal10} (see also \citealt{Righi:2008br}). For reference, $L^{\rm CO(3-2)}_0 = 7.0 \times 10^4 L_\odot$,
while nearby higher order rotational transitions have slightly higher luminosities until the luminosity declines again above CO(7-6).  The CO luminosities are based on correlations between
the strength of these emission lines and galactic star formation rates, as measured at low redshift, while the [CII] luminosity-SFR relation is normalized to M82.
Using the SFR functions from \citet{Smit12} (adopting their nearest redshift bin for each interloper redshift), we can then estimate the emissivity and average specific intensity according to 
Eqs.~\ref{eq:sfr_func}--\ref{eq:emiss}. This gives $\avg{I_t} = 5.7 \times 10^2$ Jy/str for [CII] emission at $z_t=7$. Likewise, summing over 
interloper transitions, we find a combined average interloper intensity of $\avg{I_{\rm j,combined}} = 7.0 \times 10^2$ Jy/str, after including all non-negligible CO lines. Interestingly, the interloper
and target contributions are comparable and so it will indeed be important to disentangle these two contributions. The top panel of Fig.~\ref{fig:interlop_imp} gives further information, quantifying 
which interloper
lines contribute most prominently to the total average intensity. According to our estimate, several distinct lines contribute significantly with the CO(4-3) at $z=0.88$, CO(5-4) at $z=1.4$, CO(6-5) at $z=1.8$, and CO(7-6) at $z=2.3$ transitions
each contributing more than $10^2$ Jy/str.  While these simple estimates provide a useful guide, we caution that they adopt simplistic assumptions about the relationship between star-formation and luminosity,
and extrapolate empirical correlations beyond the redshifts at which they have been determined. (See also the discussion in the beginning of this section.) Our results are nevertheless broadly consistent with previous estimates in \citet{Silva:2014ira}, but differ in the details of the modeling and the empirical constraints adopted. 
Given the uncertainties in the signal and interloper strengths, we aim to devise a flexible approach
for separating the interloper and target emission signals.  

We also checked the impact of interloper emission from additional fine structure lines: [CI] $610 \mu {\rm m}$ at $z=1.1$, [CI] $371 \mu {\rm m}$ at $z=2.4$, [NII] $205 \mu {\rm m}$ at $z=5.2$, and [OI] $145 \mu {\rm m}$ at $z=7.7$\footnote{The latter line is at slightly higher redshift than the target line, and so might instead be referred to as an ``extraloper'' line.}. In our model, the strongest of these lines is [CI] $371 \mu {\rm m}$ which has an average specific
intensity of $\avg{I_{{\rm CI}, 371 \mu {\rm m}}} = 54$ Jy/str, and so it contributes less than $10\%$ of the target emission. As justified further in the next paragraph, we neglect these potential interlopers in this work.

 In order to quantify the relative importance of the interloper transitions to the power spectrum of intensity fluctuations, which is ultimately the signal we are after, we need to consider more than just the average
 specific intensity.  Eqs. ~\ref{eq:pfor_rspace} and \ref{eq:ptar_rspace} imply that the relative strengths of the fluctuations depend mostly -- asides for the shifting of power in wavenumber -- on
 $P_{j} \propto \avg{b_{j}}^2 \avg{I_{j}}^2 D^2(z_j)/\left[\alpha_\parallel(z_j) \alpha_\perp^2(z_j)\right]$, with $D(z_j)$ being the linear growth factor at redshift $z_j$. This applies on large scales where shot-noise contributions are negligible.  We plot the relative strength
 of fluctuations, as characterized by this one number, in the bottom panel of Fig.~\ref{fig:interlop_imp}. In comparison to the average specific intensity, this number is enhanced for the lower $J$
 transitions because the distortion factor $1/\left(\alpha_\parallel(z_j) \alpha_\perp^2(z_j)\right)$ and the growth factor $D(z_j)$ increase towards lower redshift. For the power spectrum of fluctuations, the dominant emission
 comes from the CO(4-3) line in this model, and the fluctuations in this line are more than a factor of two larger than in the target [CII] line. Fluctuations from CO(3-2), CO(5-4), and CO(6-5) each contribute between $5-20\%$  of the total interloper fluctuations. Higher order transitions contribute less than several percent to the interloper fluctuations, and we will assume they contribute negligibly in what follows.
The same is true of the [CI], [NII], and [OI] interloper/extraloper lines discussed above, and so we neglect them as well.
We will discuss relaxing this assumption where appropriate; it is straightforward to include additional interloper lines in our calculations, but this adds additional parameters to the modeling.

In addition to the clustering term, we should also consider the shot-noise contribution to the power spectrum from the target and interloper lines. This contribution may also be
estimated from the $L-SFR$ correlation, and the observed SFR Schechter function fits.
Specifically, we expect the shot-noise from galaxies emitting in line $L$ to be (e.g. \citealt{Uzgil:2014pga}):
\beq
P_{\rm shot, L} = \frac{\avg{I_L}^2}{\phi_\star} \frac{2+\alpha}{\Gamma(2+\alpha)}.
\label{eq:pshot}
\eeq
Using the numbers from \citet{Smit12} for $\phi_\star$ and $\alpha$, we find $P_{\rm shot, t} = 2.9 \times 10^5$ Jy$^2$/str$^2$ $\left({\rm Mpc}/h\right)^3$ for the target emission line.
Summing over all of the interlopers, up to and including the CO(6-5) transition gives $P_{\rm shot, j, combined} = 1.2 \times 10^7$  Jy$^2$/str$^2$ $\left({\rm Mpc}/h\right)^3$, after including
the distortion factors. The shot noise from the interlopers is hence almost $50$ times that in the target emission. In this work we will consider the combined target plus interloper shot-noise
as a single ``nuisance'' term that we aim to subtract out.

\subsection{Apparent Interloper and Signal Power Spectra}

\begin{figure}
\begin{center}
\includegraphics[width=9cm]{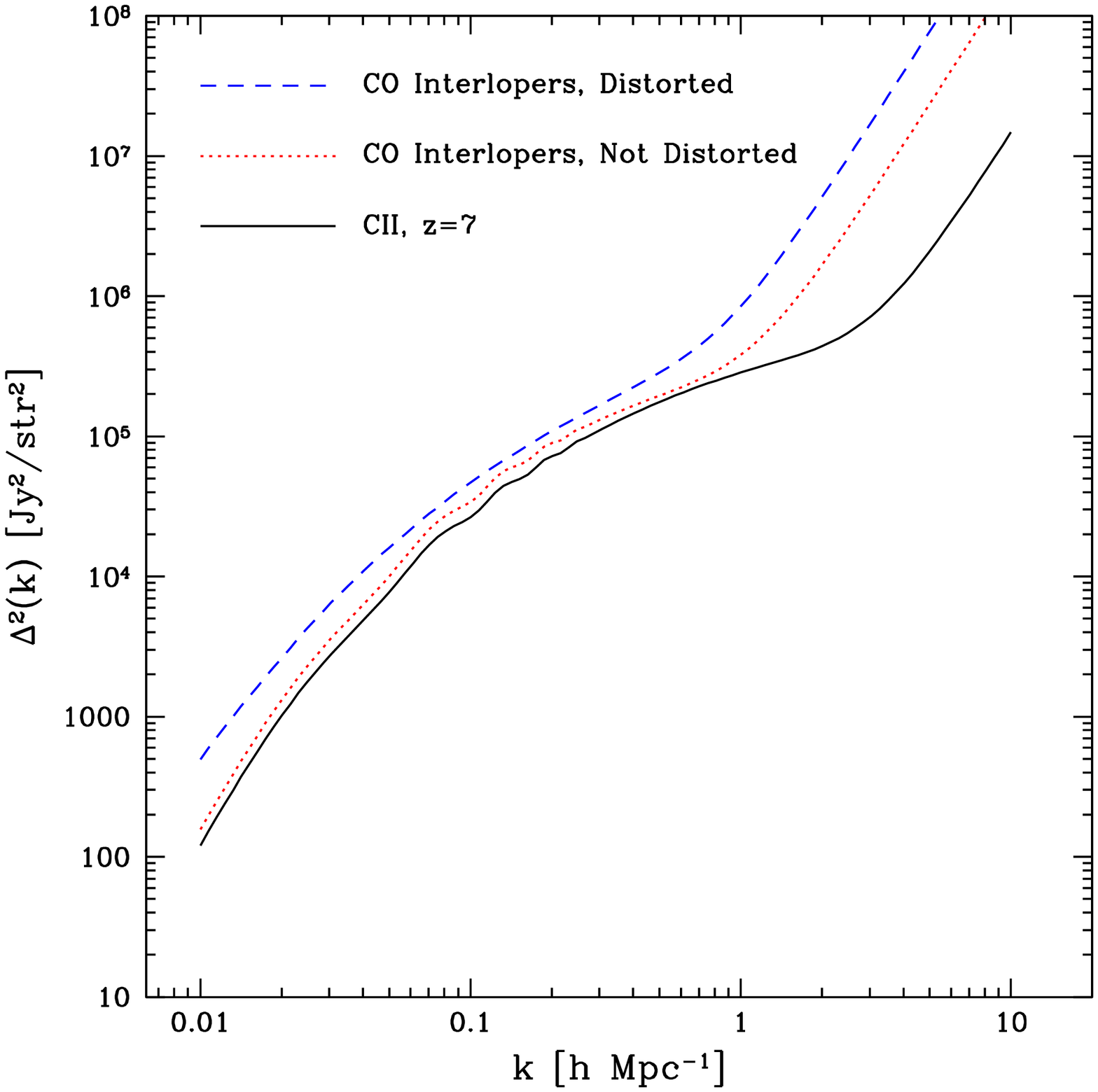}
\includegraphics[width=9cm]{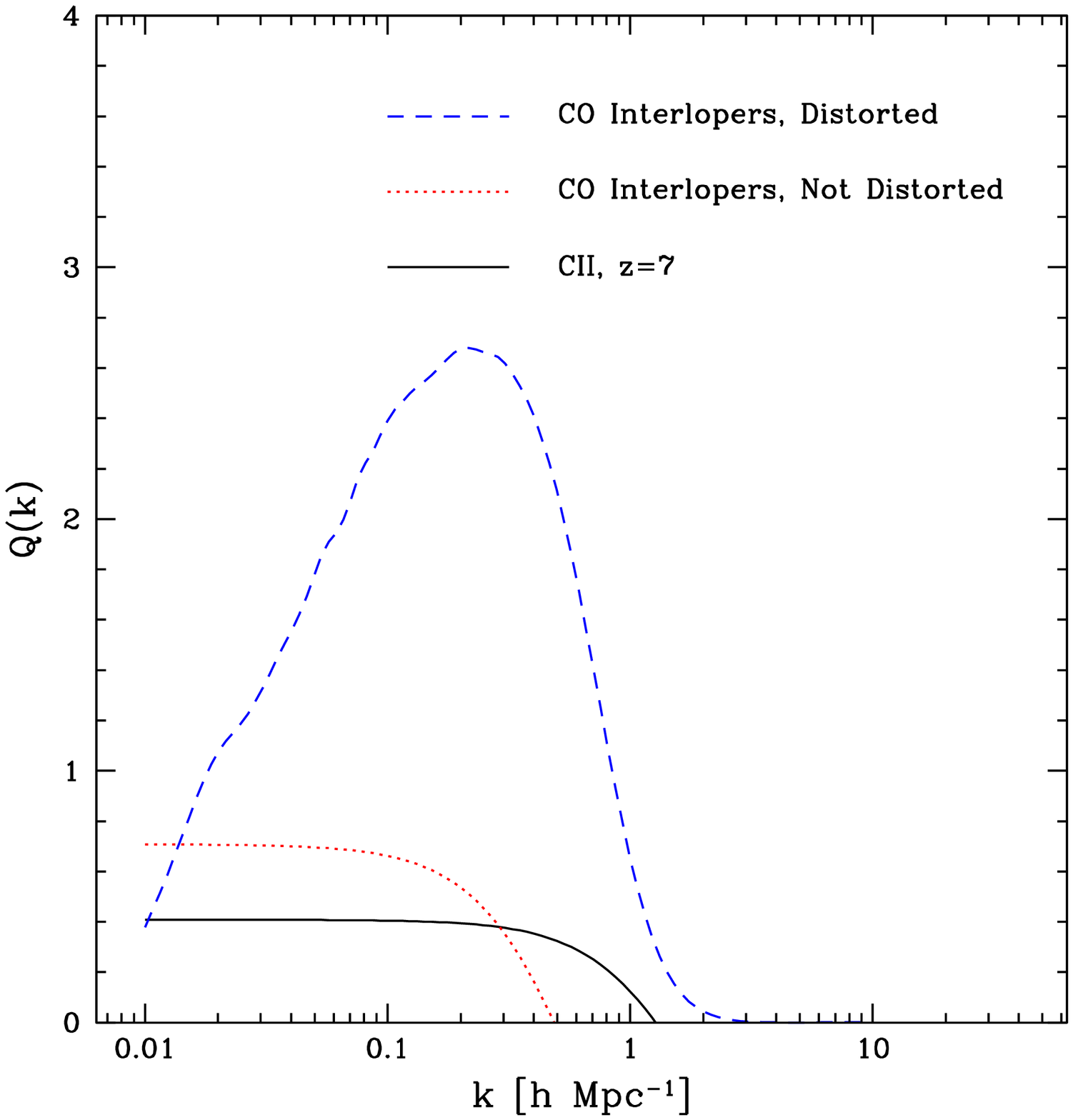}
\caption{Monople and quadropole to monopole ratios for the interloper and signal power spectra. {\em Top}: The power spectrum monopole, multiplied by $k^3/(2 \pi^2)$ so that each line shows the
usual spherically-averaged contribution to the variance per ${\rm ln}(k)$.  The lower black solid line shows our model for the target [CII] intensity fluctuation power at $z_t=7$. The blue dashed line
shows our model for the total CO$(J \rightarrow J-1)$ contamination including the impact of the coordinate distortions. The red dotted line shows the true total interloper power spectrum monopole, neglecting the remapping
effects. {\em Bottom}: The quadropole to monopole ratio of the power spectra in each case. The true power spectra are anisotropic only because of the Kaiser and finger-of-god effects, while incorporating the
remapping distortions boosts the interloper quadropole to monopole ratio on large scales. These ratios turn over on small scales (high $k$) due to the finger-of-god
effect and shot noise contamination.}
\label{fig:pkmon_quad}
\end{center}
\end{figure}

We now turn to examine the model signal and interloper power spectra. As a first convenient way of characterizing the target and interloper power spectra, we expand the spectra
in terms of Legendre polynomials and calculate the monopole and quadropole moments. 
The quadropole-to-monopole ratio may be written as
\beq
Q(k) =  \frac{\frac{5}{2} \int_{-1}^{1} d\mu \left[3\mu^2/2 - 1/2\right] P(k,\mu)}{\frac{1}{2} \int_{-1}^{1} d\mu P(k,\mu)}.
\label{eq:quad_to_mon}
\eeq
We can calculate the intrinsic target quadropole to monopole ratio, as well as that for the apparent interloper power spectra, incorporating the distortions as described by
Eqs.~\ref{eq:power_warp}--\ref{eq:pfor_rspace}. 

The spherically averaged (monopole) power spectra are shown in the top panel of Fig. \ref{fig:pkmon_quad}. The solid black line shows the target [CII] emission power spectrum at
$z_t=7$, $k^3 P_t(k)/(2 \pi^2)$. In this model, the [CII] power spectrum has a strength of about $\Delta^2 \approx 10^2$ Jy$^2$/str$^2$
at $k \sim 0.01 h$ Mpc$^{-1}$, $\Delta^2 \approx 3 \times 10^5$ Jy$^2$/str$^2$ at $k \sim 1 h$ Mpc$^{-1}$, and reaches $\Delta^2 \approx 1.5 \times 10^7$ Jy$^2$/str$^2$ at $k \sim 10 h$ Mpc$^{-1}$. The clustering
term dominates on large scales at $k \lesssim 3 h$ Mpc$^{-1}$ or so, while the shot-noise term is more important on smaller scales. The blue-dashed and red-dotted lines show
the interloper contamination power, with and without coordinate distortions, respectively. For each interloper line, the coordinate distortions shift power from $k_\parallel \rightarrow k_\parallel/\alpha_\parallel(z_j)$ and from $\k_\perp \rightarrow \k_\perp/\alpha_\perp(z_j)$, while boosting the fluctuation power by the overall $1/(\alpha_\parallel(z_j) \alpha_\perp^2(z_j))$ factor. After spherical averaging, this leads to a shift and boost in the apparent interloper power, as may be discerned by comparing the blue dashed and red dotted lines in Fig.~\ref{fig:pkmon_quad}. As anticipated in the previous section, the combined CO interloper power exceeds the target [CII] emission power by a factor of several on large scales -- the precise excess depends on scale because of the coordinate distortions -- and so it is crucial
to remove this contamination. On smaller scales the target and interloper power differ because of the larger Poisson noise from the interloper populations: we expect the interloper shot-noise
to swamp that in the [CII] target emission. The larger interloper shot-noise mostly results because star-formation occurs in lower mass, yet more abundant systems at high redshift and so the Poisson noise
in the high redshift target line is relatively low.
As mentioned previously, in this work we will be content to extract only the [CII] clustering term and forego trying to separate out the [CII] shot-noise term in the presence of this
large interloper contamination.

Although the shape of the target and interloper monopole power differ only subtly, the angular dependence of the target and interloper power is quite different. For example, the bottom panel
of Fig. \ref{fig:pkmon_quad} shows the quadropole to monopole ratio for both the target and interloper emission power spectra. For illustration, we show the CO interloper quadropole to monopole
ratio both with and without coordinate mapping distortions. The quadropole to monopole ratio for the target emission, and the interloper emission without coordinate mapping distortions, have the usual
form expected from redshift space distortions. On sufficiently large scales, $Q \rightarrow (4 \beta/3 + 4 \beta^2/7)/(1 + 2 \beta/3 + \beta^2/5)$ -- the  Kaiser effect result \citep{Kaiser87} --  
while the quadropole anisotropy diminishes on smaller scales owing to the finger-of-god effect and the isotropic shot-noise term. The intrinsic interloper $Q(k)$ turns over on larger scales (smaller $k$) than the target $Q(k)$ because the interloper shot-noise term is bigger and because the finger-of-god suppression is stronger at the (lower) redshifts of the interloper lines. The blue dashed line shows the quadropole to monopole ratio after incorporating the coordinate
mapping distortion. This {\em reaches much larger values} than expected from the Kaiser effect, with the model $Q$ peaking near $Q=2.6$ at $k=0.2 h$ Mpc$^{-1}$ before gradually turning over on smaller
scales owing to the finger-of-god effect and shot-noise.
 This is a direct consequence of the difference between the mapping factors, $\alpha_\parallel(z_j)$ and $\alpha_\perp(z_j)$, and the shape of the linear power spectrum of density fluctuations. The increasing $Q(k)$ from $k \sim 0.01-1 h$ Mpc$^{-1}$ reflects the steepening of the power spectrum
spectral index towards small scales. This can be verified by calculating the quadropole to monopole ratio for a pure power law power spectrum (of varying spectral index) 
under the coordinate warping transformation.
The steeper $k$ dependence at small scales enhances the difference between the line of sight and transverse power after applying the warping.  Note that on scales larger than the co-moving horizon
size at matter radiation equality, $k \leq k_{\rm eq} \sim 0.015 h$ Mpc$^{-1}$, the net interloper distortion is sub-Kaiser because the linear matter power spectrum is an increasing function of $k$ on
these scales.

\begin{figure}
\begin{center}
\includegraphics[width=0.45\textwidth]{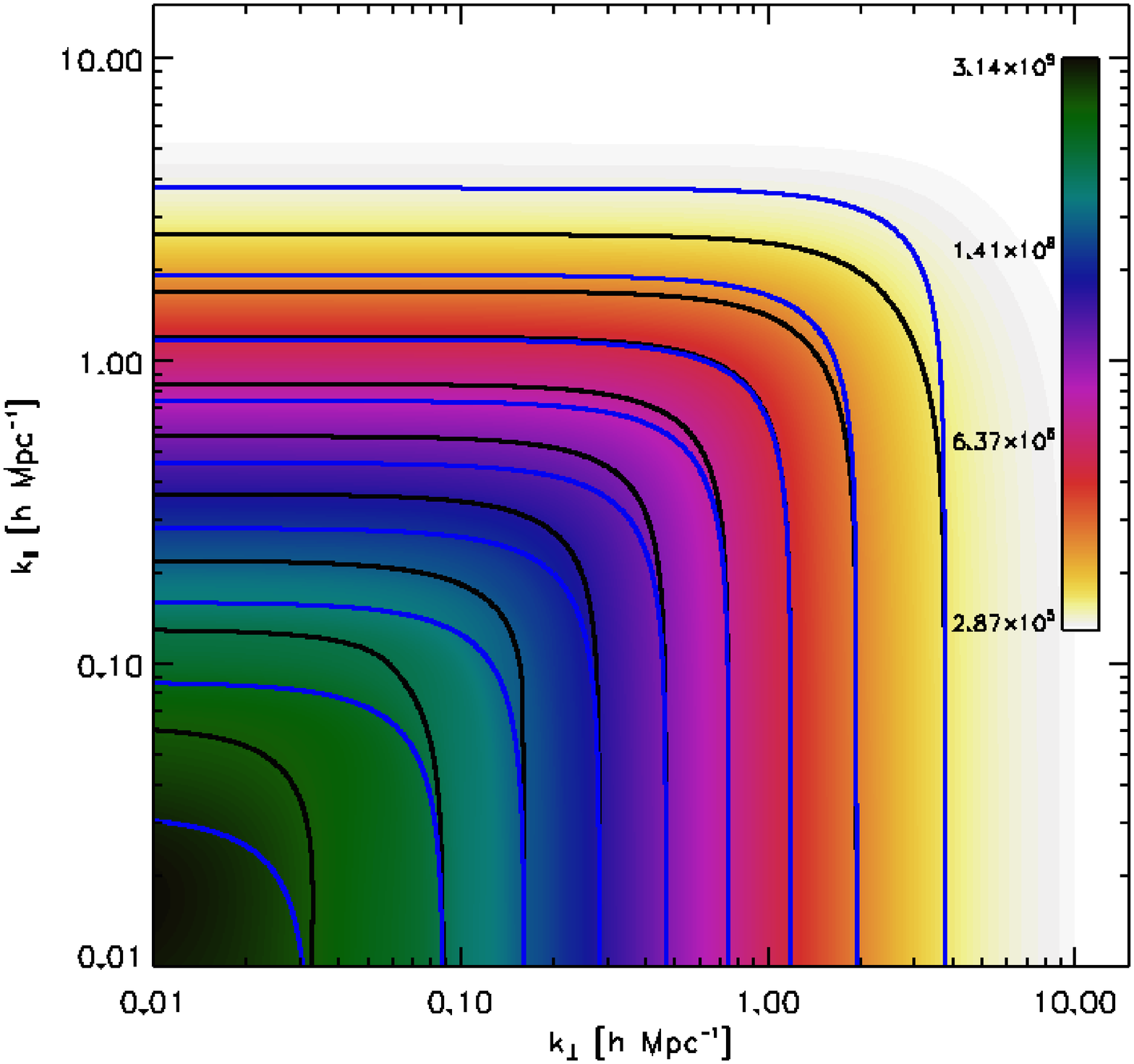}
\includegraphics[width=0.45\textwidth]{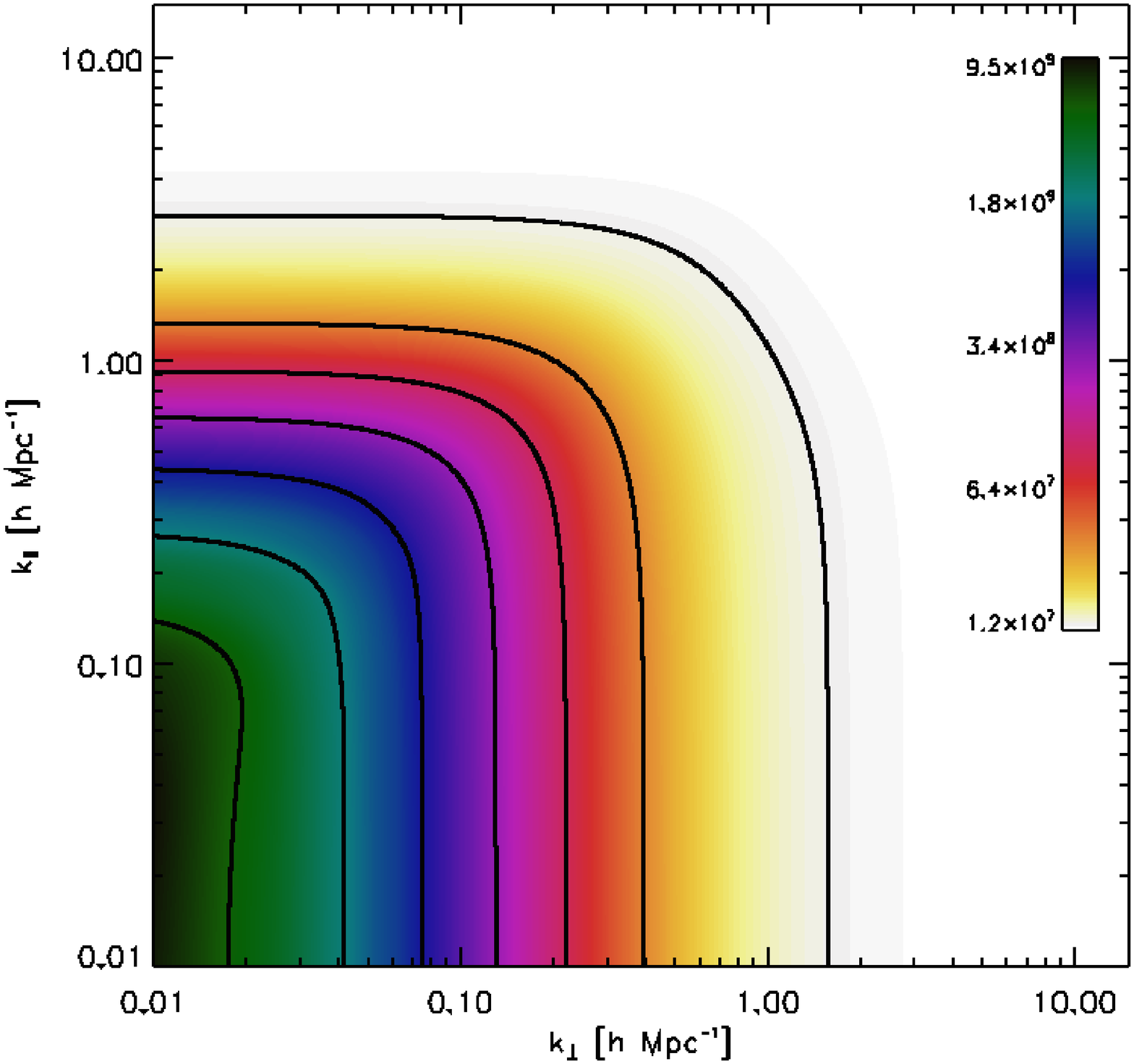}
 \caption{Anisotropy of the target and interloper power spectra from redshift space and coordinate mapping distortions. {\em Top}: Contours of constant power in the $k_\perp-k_\parallel$ plane for
 the target [CII] power spectrum at $z_t=7$. The blue contours neglect redshift space distortions, while the black contours and color-scale incorporate them. On large scales, the power spectra show the Kaiser enhancement
 for wave numbers in the line of sight direction. This effect turns around and the contour just past $k \sim 1 h$ Mpc$^{-1}$ is nearly isotropic, with the finger-of-god effect dominating at slightly higher
 $k$ until isotropic shot-noise dominates. The colorbar is in units of $({\rm Jy/str})^2 ({\rm Mpc}/h)^3$.
 The lowest contour is at $P(k) = 5 \times 10^5 ({\rm Jy/str})^2 ({\rm Mpc}/h)^3$ and the contours increase inwards as $d{\rm ln}P=1$.  {\em Bottom}: The black contours and color-scale show the anisotropy of the total interloper emission power spectrum including the coordinate mapping distortion. The lowest contour is at $P(k) = 1.3 \times 10^7 ({\rm Jy/str})^2 ({\rm Mpc}/h)^3$ and the contours increase inwards as $d{\rm ln}P=1$. The interloper anisotropy is much stronger than that in the target line; this can be used to separate-out the interloper contamination.}
 \label{fig:contour_aniso}
 \end{center}
 \end{figure}

In order to further characterize and visualize the target and interloper power spectrum anisotropies, we plot contours of constant power in the $k_\perp-k_\parallel$ plane (Fig.~\ref{fig:contour_aniso}, see also \citealt{Gong:2013xda}). The top panel illustrates the redshift space distortion in the target emission. As mentioned earlier, if the large scale anisotropy shown here can be measured accurately, we can determine the luminosity-weighted bias of the emitters in the target line (from the dependence
on $\beta_t$), as well as the average specific intensity of the target emission (from the overall amplitude of fluctuations). The contours in the bottom panel show a strong elongation in the $k_\parallel$
direction from the coordinate mapping distortion, which sources the strong quadrupole moment shown in Fig.~\ref{fig:pkmon_quad} as discussed previously.
Note that the total interloper power spectrum in our model is the sum of four separate interloper lines from different redshifts, CO(3-2) at $z=0.407$, CO(4-3) at $z=0.877$, CO(5-4) at
$z=1.35$, CO(6-5) at $z=1.82$. The total anisotropy of the interloper emission, illustrated in Fig.~\ref{fig:pkmon_quad} and Fig.~\ref{fig:contour_aniso} hence reflects a weighted average of these
four interloper lines, with the CO(4-3) line having the strongest weight in our model (see the bottom panel of Fig.~\ref{fig:interlop_imp}).

\section{Forecasts} \label{sec:forecasts}

Having quantified the power spectrum anisotropy, we now forecast the prospects for using this to separate out the interloper and target contributions
to the power spectrum. Here we assume that
 Eqs.~\ref{eq:power_warp_tot}--\ref{eq:pfor_rspace} provide a perfect description of the measured power spectra. We then investigate how well the parameters of the model may be determined by hypothetical [CII] surveys. The shortcoming of this approach
 is that it relies on simple models for the power spectra of intensity fluctuations, which may be imperfect. In future work, it will be important to develop consistency checks of this model, and/or
 to develop a more sophisticated description. We discuss some possible observational tests in \S \ref{sec:cross} and \S \ref{sec:cross_lines}. 
 
In general, we consider a seven-dimensional parameter space described
by a vector, $\q$, with seven components: $\{q_1, q_2, ....., q_7\} = \{\avg{I_t}, \avg{b_t}, \avg{I_{32}},\avg{I_{43}},\avg{I_{54}},\avg{I_{65}},P_{\rm shot, tot}\}$. The parameters describe the specific
intensity of the target emission, the average bias of this emission, the specific intensity of each of the four important interloper lines (indexed by the rotational states of the CO transitions with $J,J-1$ as subscripts: e.g., $\avg{I_{32}}$ is the average specific intensity in the $J=3 \rightarrow 2$ transition), and the total (target plus all interlopers) shot-noise.
Here we implicitly fix the bias of the fluctuations in each interloper line to $\avg{b_j} = 2$. Since the interloper power is determined mostly by the overall product of specific intensity and bias (asides
for the additional dependence on $\beta$ through the Kaiser effect which is small relative to the anisotropy induced by assuming an incorrect redshift), one can think of the specific intensity
constraints derived as confidence intervals on the product $\avg{b_j} \avg{I_j}$. 

Our main goal then is to determine whether the target emission fluctuations, characterized by the parameters $\avg{I_t}$ and $\avg{b_t}$, may be determined accurately in the presence of the interloper fluctuations. We investigate this by calculating Fisher matrices for futuristic [CII] surveys.
The components of the Fisher matrix for parameters $q_i$ and $q_j$ are given by:
\begin{align}
F_{ij} =& \int_{\mu_{\rm min}}^{\mu_{\rm max}} d \mu \nonumber \\
& \times  \int_{k_{\rm min}}^{k_{\rm max}} \frac{dk k^2 V_s}{4 \pi^2}  \frac{\partial P(k,\mu)}{\partial q_i} \frac{\partial P(k,\mu)}{\partial q_j} \frac{1}{{\rm var}[P(k,\mu)]},
\label{eq:fisher_ij}
\end{align}
where we have approximated the discrete sum over modes in the survey by a continuous integral.
Here the integral over angle runs over the upper half-plane, between some ($k$-dependent) limits $\mu_{\rm min}$ and $\mu_{\rm max}$ that we will describe below, and the integral over wavenumber
ranges between the limits $k_{\rm min}$ and $k_{\rm max}$. The quantity $V_s$ is the co-moving volume of the survey. This expression depends on that variance of the total power spectrum of fluctuations for each $\k$-mode, ${\rm var}[P(k,\mu)]$. We compute this, neglecting non-Gaussian contributions to the variance, as:
\beq
{\rm var}[P(k,\mu)] = \left[P_{\rm tot}(k,\mu) + P_N(k,\mu)\right]^2.
\label{eq:varpk}
\eeq
Here $P_{\rm tot}(k,\mu)$ is the total signal plus interloper emission power spectrum, including the shot-noise contribution, and $P_N(k,\mu)$ is the detector noise power spectrum. 

It is also instructive to consider the number of Fourier modes in the upper-half plane in a bin of $k$ and $\mu$, $N_m(k)$. For a survey of co-moving volume $V_s$, the number of modes contained within the survey volume in a wavenumber bin of thickness $\Delta {\rm ln}(k) \Delta \mu$ is:
\beq
N_m(k) = \frac{k^3 V_s}{4 \pi^2} \Delta {\rm ln}(k) \Delta \mu.
\label{eq:nmode}
\eeq
Note that this is just included for illustration, since the mode-counting is already handled implicitly in the Fisher matrix calculation (Eq.~\ref{eq:fisher_ij}).

\subsection{Survey Parameters}

It will be challenging to measure the power spectrum and its angular dependence precisely enough to separate the faint interloper and target signals using this methodology. Nevertheless, experiments are already underway to detect the reionization-era [CII] signal (e.g. the TIME-Pilot experiment, \citealt{Crites14}); we anticipate that the sensitivity of  these measurements will increase rapidly, fueled by
advances in detector technology. As a convenient baseline, we consider the ``CII-Stage II'' survey described in \citet{Silva:2014ira}. Unfortunately, we find that even this is less sensitive than we
require and so we generally consider a still more sensitive experiment, as specified subsequently. 

Our baseline survey is described in \citet{Silva:2014ira} and consists of a single $10$ meter dish, with $16,000$ bolometers and an $N_{\rm sp}=64$-beam spectrometer with a frequency resolution
of $\Delta \nu = 0.4$ GHz.  The hypothetical survey spans $100$ deg$^2$ on the sky for a total observing time of $t_{\rm survey} = 2,000$ hours. 
We consider a $B=20$ GHz bandwidth of observations near $z=7$, which is small enough for us to neglect evolution in the signal across the survey bandwidth. The angular resolution of
the survey is $\Delta \theta = 0.43$ arcminutes.  In co-moving coordinates, the pixels span $x_{\perp,res} = 0.790$ Mpc/$h$ in the transverse direction
and $x_{\parallel,res} = 3.42$ Mpc/$h$ in the line of sight direction.  In the line of sight direction, the survey length is $L_\parallel = 171$ Mpc/$h$, while the transverse dimension is $L_\perp = 1.09 \times 10^3$ Mpc/$h$. The total survey volume is $V_s = 2 \times 10^8 ({\rm Mpc}/h)^3$. 
For reference, the number of modes surveyed is
$N_m(k) = 5.2 \times 10^3 (k/0.1 h {\rm Mpc}^{-1})^3 \Delta {\rm ln}(k) \Delta \mu$ in a bin around $k=0.1 h$ Mpc$^{-1}$. 

The survey noise power spectrum may be written as (e.g. \citealt{Uzgil:2014pga}):
\beq
P_N(k_\parallel,k_\perp) = \frac{\sigma_N^2}{t_{\rm obs}} V_{\rm pix} e^{(k_\parallel x_{\parallel,res})^2 + (k_\perp x_{\perp,res})^2},
\label{eq:pnoise}
\eeq
where $\sigma_N^2/t_{\rm obs}$ is the noise per pixel in specific intensity units (squared), $V_{\rm pix}$ is the pixel volume, and the exponential factor accounts for the finite angular and spectral resolution of
the instrument.  We can extract plausible numbers for the noise power spectrum from Table 8 of \citet{Silva:2014ira}, converting from the Noise Equivalent Flux Density (NEFD) to the 
specific intensity noise($\times$ square-root
of time in seconds), using $\sigma_N = {\rm NEFD}/(\Delta \Omega_{\rm pix})$. Note also  
that the observing time per pixel is $t_{\rm obs} = t_{\rm survey} N_{\rm sp} \Delta \Omega_{\rm pix}/ \Delta \Omega_{\rm survey}$, where $N_{\rm sp}$ is the number of spatial pixels and
$t_{\rm survey}$ is the total survey observing time.

This resulting noise power spectrum is:
\begin{align}
\frac{\sigma_N^2}{t_{\rm obs}} V_{\rm pix} =& 8.7 \times 10^8 \frac{{\rm Jy}^2}{{\rm str}^2}\left(\frac{{\rm Mpc}}{h}\right)^3 \left[\frac{\sigma_N}{3.1 \times 10^5 {\rm Jy/str} \sqrt{{\rm sec}}}\right]^2 \nonumber \\
& \times \left[\frac{V_{\rm pix}}{2.13 ({\rm Mpc}/h)^3}\right] \left[\frac{64}{N_{\rm sp}}\right] \left[\frac{2,000 {\rm hrs}}{t_{\rm survey}}\right].
\label{eq:pn_hyp}
\end{align}
Since we find that even this sensitivity is insufficient for our purposes, we consider a still more sensitive experiment with $(\sigma_N^2 V_{\rm pix})/t_{\rm obs} = 4.3 \times 10^7 {\rm Jy}^2 {\rm str}^{-2}
({\rm Mpc}/h)^3$. This value represents our fiducial noise level in what follows. We caution that the noise power here is approximately twenty times smaller than in the Stage-II experiment considered by \citet{Silva:2014ira}, and so the rms noise in our fiducial case is $4-5$ times smaller than in this previous work. Naturally, it will be important to see if this sensitivity is in fact achievable.
Improvements may be possible by going to space, in which case the CMB would set the photon background noise rather than emission from the Earth's atmosphere. Rapid progress in detector development
may also help to increase sensitivity beyond what is assumed here, e.g. it may be possible to increase the number of spatial pixels, $N_{\rm sp}$.
 We will describe
how the results depend on this somewhat arbitrary choice of noise power. It may also be possible to make progress with noisier survey data by masking bright pixels suspected of containing CO interloper emission, while
using the anisotropy of the residual fluctuation power spectrum to further clean interloper contamination. In other words, the masking approach advocated in previous work may be combined
with the technique developed here. Finally, there may be some benefit to a spare-sampling survey strategy to build up a large field-of-view quickly -- rather than mapping contiguous regions on the sky -- although this will lead to aliasing from high-$k$ modes (\citealt{Kaiser:1996tp,Chiang:2013ksa}). 

We are almost ready to calculate the Fisher matrix elements using Eq.~\ref{eq:fisher_ij}, but we need to comment first on one additional complication. The issue relates to the continuum foreground, which is significantly larger than the line interloper emission. The continuum emission at the frequencies of interest is dominated by the Cosmic Infrared Background (CIB), produced by dust grains in galaxies
at a range of redshifts, and has an average specific intensity of a $\sim$ a few $\times 10^5$ Jy/str \citep{Silva:2014ira}. Although this is two to three orders of magnitude larger than the expected [CII] emission, the continuum foreground should nevertheless be separable using the fact that it is spectrally smooth, i.e., one can use exactly the same strategy as advocated for cleaning foregrounds from 
redshifted 21 cm fluctuation measurements (e.g \citealt{Zaldarriaga:2003du}). In order to separate the spectrally smooth foreground, however, one inevitably sacrifices measuring long wavelength modes along the line of sight. Additional modes will likely be lost as well, since the frequency dependence of
the beam, calibration errors, and
instrument imperfections can also produce spurious spectral structure in the foregrounds, {\em as observed by the instrument}. Here we will ignore this ``mode-mixing'' problem (e.g. \citealt{Liu11,Ali15}), and take a simplistic approach: we simply remove line-of-sight modes
with wavelength smaller than the bandwidth of the measurement, i.e. modes with line-of-sight wavenumber smaller than
$k_{\parallel, {\rm min}} = 2 \pi/L_\parallel = 0.037 h$ Mpc$^{-1}$.  Further work is required to determine whether measuring the angular dependence of the power spectrum is feasible in the presence
of realistic levels of mode-mixing. Mode-mixing should be significantly less bad here than in the case of 21 cm; in part this is because the continuum to line emission ratio is smaller, and also because the
instrumental beam is simpler for this single dish experiment.

\begin{figure}
\begin{center}
\includegraphics[width=9cm]{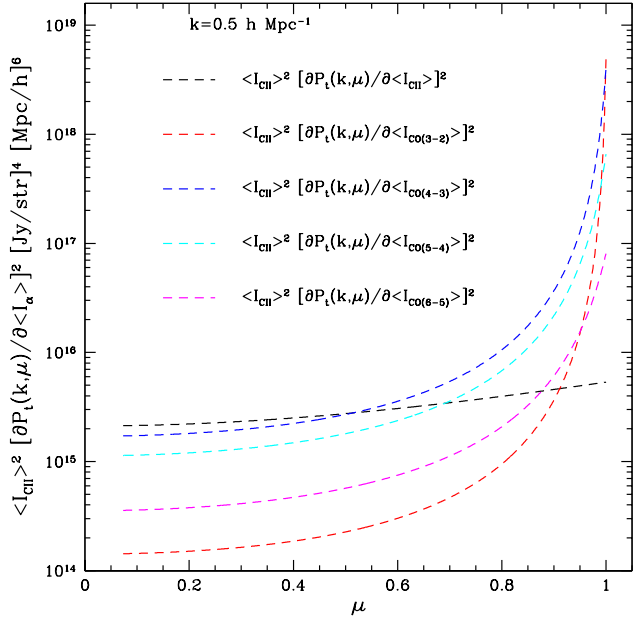}
\caption{Derivatives with respect to average specific intensity in different lines as a function of angle. Here the power spectrum derivatives are computed for $k=0.5 h$ Mpc$^{-1}$.  The derivatives 
are much stronger functions of angle for the interloper line -- owing to the coordinate distortions -- than for the target [CII] emission, which depends on angle only through the Kaiser effect. The
low order CO transitions show a stronger dependence on angle because their coordinate distortions are larger.}
\label{fig:fisher_examp}
\end{center}
\end{figure}

Before exploring forecasts for marginalized constraints on the parameters $\q$, it is instructive to explicitly examine some of the derivatives that enter the Fisher matrix calculation of Eq.~\ref{eq:fisher_ij}. Fig.~\ref{fig:fisher_examp} compares the derivatives of the total power spectrum with respect to each of the specific intensity parameters as a function of angle, $\mu$, for fixed $k = 0.5 h$ Mpc$^{-1}$. The range 
of $\mu$
accessible is limited slightly by removing the spectrally smooth modes with $k_{\parallel} \leq 2 \pi/L_\parallel$ to $\mu \geq \mu_{\rm min} = k_{\parallel, {\rm min}}/k = 0.074$ for $k=0.5 h$ Mpc$^{-1}$. 
The derivatives with respect to the interloper line intensities show a steeper
angular dependence than the target emission, as expected. This is because the interlopers are subject to the coordinate distortion, while the target line depends on angle primarily through the Kaiser effect (at the wavenumber considered, the finger of god effect is sub-dominant). A simple way to understand the angular dependence of the interloper power is to note that, approximating the power spectrum at the wavenumber of interest by a power-law of spectral index $k^{-n_{\rm eff}}$, the ratio of the power at $\mu=1$ to that at $\mu=0$ is simply $(\alpha_\parallel(z_j)/\alpha_\perp(z_j))^{n_{\rm eff}}$. This rough
estimate ignores the Kaiser effect, which will further enhance this ratio. For $k=0.5$ h Mpc$^{-1}$, $n_{\rm eff}  = -d{\rm ln}P/d{\rm ln} k = 2.1$\footnote{A more detailed estimate would also take into account that the local spectral index should really be evaluated separately at each of $k/\alpha_\parallel$ and $k/\alpha_\perp$.} , and this ratio is $110$ for the strongest case of the CO(3-2) interloper distortion. The derivative shown is proportional to the square of this number and so the ratio reaches four orders of magnitude, and the result in Fig.~\ref{fig:fisher_examp} is still slightly larger because it includes the
Kaiser distortion. 

In any case, Fig.~\ref{fig:fisher_examp} further motivates that the angular dependence can be used to separate the target and interloper contributions to the power spectrum if it can be measured with small enough error bars. In addition, comparing the angular dependence of the derivatives with respect to the intensity in the various lines gives some sense for which lines will be most degenerate with each
other. For example, the weaker angular dependence of the target line derivative suggests that $\avg{I_{\rm CII}}$ should not be strongly degenerate with the intensities in the interloper lines, provided the full angular range shown is well-measured. This should be  
especially so in comparison to the low-order transitions that show the strongest angular variation. On the other hand, we expect the intensity in the CO(4-3) and CO(5-4) lines to be more degenerate given their relatively similar redshifts and distortion factors.

\begin{figure}
\begin{center}
\includegraphics[width=0.45\textwidth]{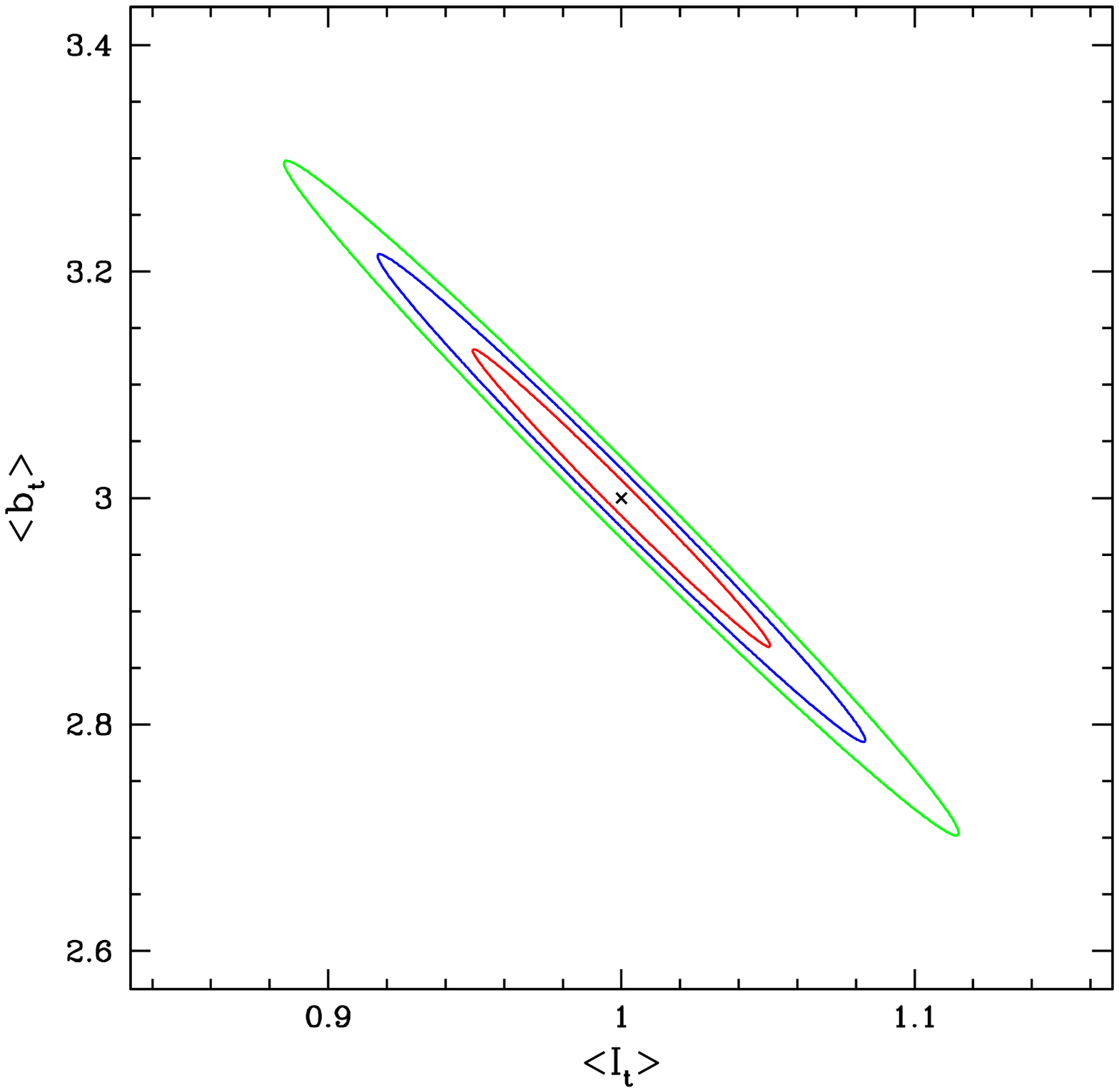}
\includegraphics[width=0.45\textwidth]{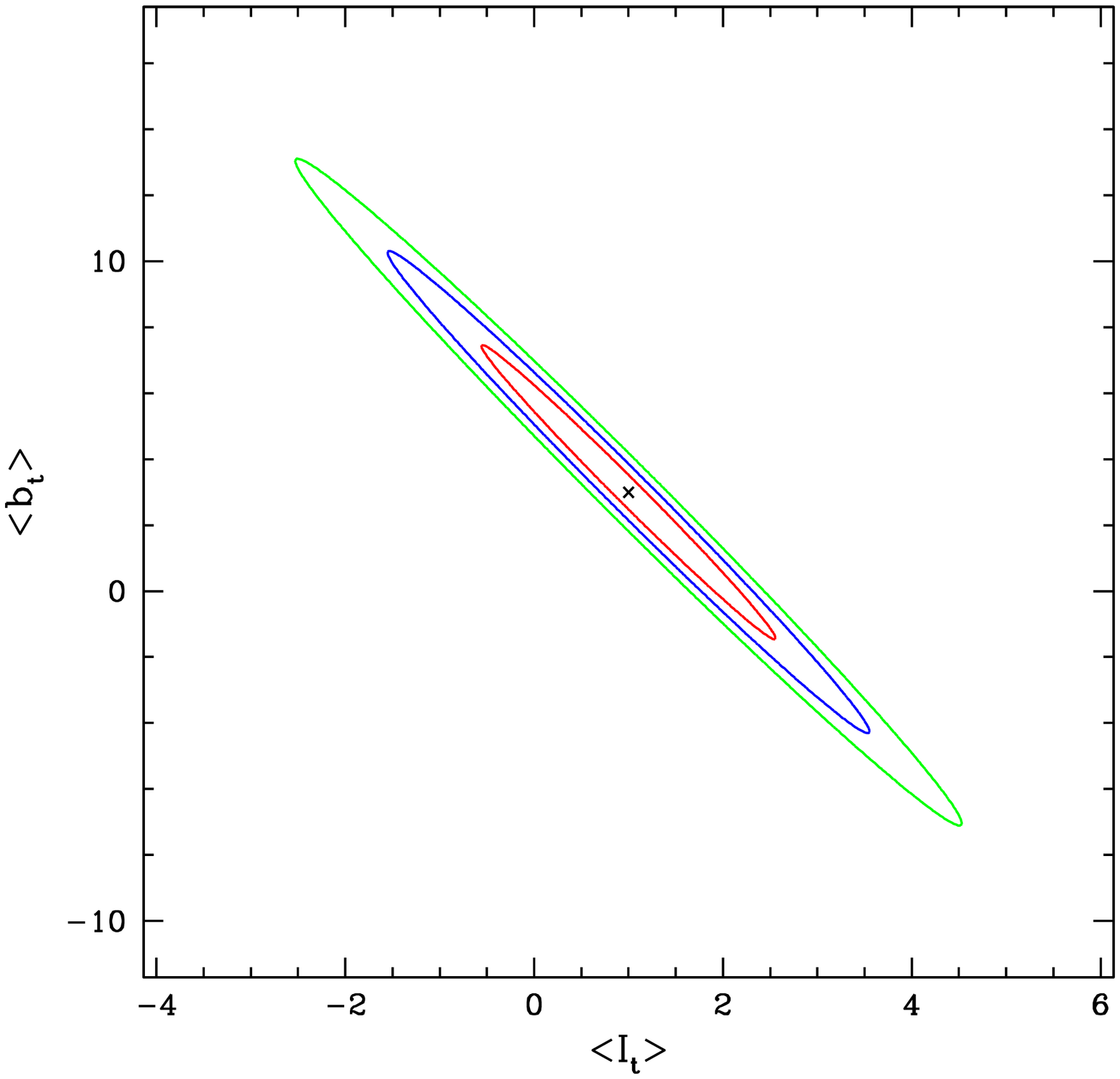}
\caption{Forecast of constraints in the $\avg{I_t}-\avg{b_t}$ plane, marginalized over the interloper contamination and shot-noise parameters for two noise levels. {\em Note that the range of values shown along
the x and y-axes differ significantly between the two panels.}  
The red, blue, and green contours show $1, 2$, and $3-\sigma$ confidence intervals, while the ``x'' marks the assumed central value.  (The contours and ``x''s have the same meaning in subsequent plots.) The specific intensity, $\avg{I_t}$, has been expressed in units of the our fiducial model value, $\avg{I_t} = 5.7 \times 10^2$ Jy/str. {\em Top:} In this case, a sample-variance limited experiment is shown, i.e., the noise power spectrum is taken
to be negligibly small. {\em Bottom:} Here the noise power spectrum instead follows Eq.~\ref{eq:pn_hyp}, as expected for the ``stage-II'' [CII] survey. Evidently, the target fluctuations can be extracted using
the angular dependence of the emission power spectrum but greater sensitivity is required than in the hypothetical stage-II survey.}
\label{fig:intensity_bias_constraints_pnoise}
\end{center}
\end{figure}

We now turn to calculate the Fisher matrix of Eq.~\ref{eq:fisher_ij}, and invert this matrix to find the constraints on the various parameters. We first consider the constraints on the target
signal, contrasting the results for the stage-II survey with noise power at the level of Eq.~\ref{eq:pn_hyp} and a sample-variance (also known as ``cosmic-variance'') limited survey, with negligibly small noise power, over the
same volume. The top panel of Fig.~\ref{fig:intensity_bias_constraints_pnoise} shows the projected errors in the $\avg{I_t}-\avg{b_t}$ plane, marginalized over the interloper parameters.
The contours show that the hypothetical sample-variance limited survey is capable of constraining $\avg{I_t}$ and $\avg{b_t}$, even in the presence of strong interloper contamination.
Quantitatively, we forecast $\sim 3\%$ level $1-\sigma$ marginalized constraints on these parameters in the sample variance limit. The ellipse
shows the expected strong degeneracy between increasing $\avg{I_t}$ and decreasing $\avg{b_t}$; nevertheless, the Kaiser effect allows separate constraints on the two parameters although they are highly correlated. However, it is hard to achieve the requisite sensitivity given the bright night sky at these frequencies. If we instead incorporate noise at the level of Eq.~\ref{eq:pn_hyp}, the marginalized errors
blow up considerably, as illustrated by the bottom panel of Fig.~\ref{fig:intensity_bias_constraints_pnoise}. In this case the marginalized constraints on the average specific intensity and the bias
only give $1-\sigma$ detections -- i.e., without attempting to mask interloper emission, a significant detection is not possible for this survey.

\begin{figure}
\begin{center}
\includegraphics[width=9cm]{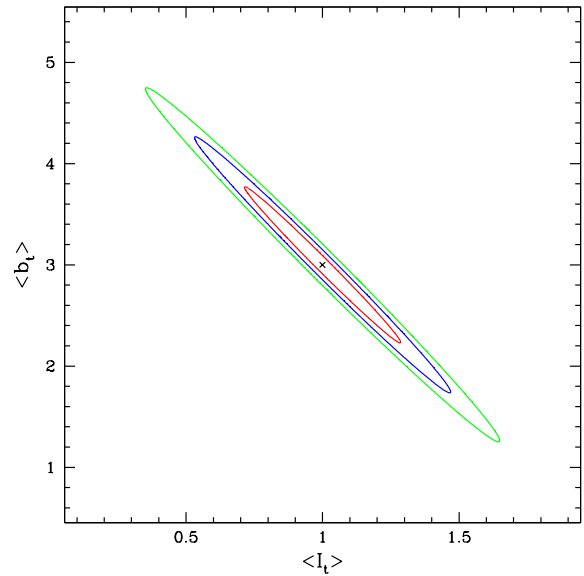}
\caption{Forecast of constraints in the $\avg{I_t}-\avg{b_t}$ plane, marginalized over the interloper contamination and shot-noise parameters for our fiducial noise power spectrum. Identical to
Fig.~\ref{fig:intensity_bias_constraints_pnoise}, except for our fiducial noise level (see text).  
}
\label{fig:intensity_bias_constraints}
\end{center}
\end{figure}

For now, we simply consider a more sensitive experiment with $(\sigma_N^2 V_{\rm pix})/t_{\rm obs} = 4.3 \times 10^7 {\rm Jy}^2 {\rm str}^{-2}({\rm Mpc}/h)^3$. The 
$\avg{I_t}-\avg{b_t}$ results, marginalized over the interloper parameters, are shown for this level of noise in Fig.~\ref{fig:intensity_bias_constraints}. Unless otherwise noted, we adopt this value
for the noise power spectrum in what follows. In this case, $20\%$ level constraints on the target emission parameters are achievable (at $1-\sigma$) and the target and interloper emission
fluctuations can indeed be separated.

The [CII] emission signal at $z \sim 7$ may also be stronger than in the model considered here, which could relax the stringent requirements on the noise power spectrum found here. Indeed, as
we were finalizing this manuscript we learned of similar work by Cheng \& Chang (2016, in prep).\footnote{Thanks to the ``Opportunities and Challenges in Intensity Mapping Workshop'' held at Stanford.} These
authors' model gives a $z=6$ [CII] emission signal that is more than an order of magnitude larger than our $z=7$ predictions, and so they are more optimistic about the prospects of applying this
method using upcoming datasets. For the most part, the difference stems from the larger bias factor in their model, with $\avg{b_t}^2$ almost six times as large as in our calculations. Their bias
factor comes from relating the line luminosity to the CIB and from empirically-calibrated models connecting CIB luminosity and halo mass. Since most of the CIB emission comes from lower redshift,
the $z=6-7$ predictions are still, however, uncertain. In any case,  this further illustrates the uncertainties in forecasting the expected signal. Improved constraints on the relationships between line-luminosity, star formation rate, and halo mass, will be needed to refine our predictions for the target and interloper emission fluctuations.

\begin{figure}
\begin{center}
\includegraphics[width=0.45\textwidth]{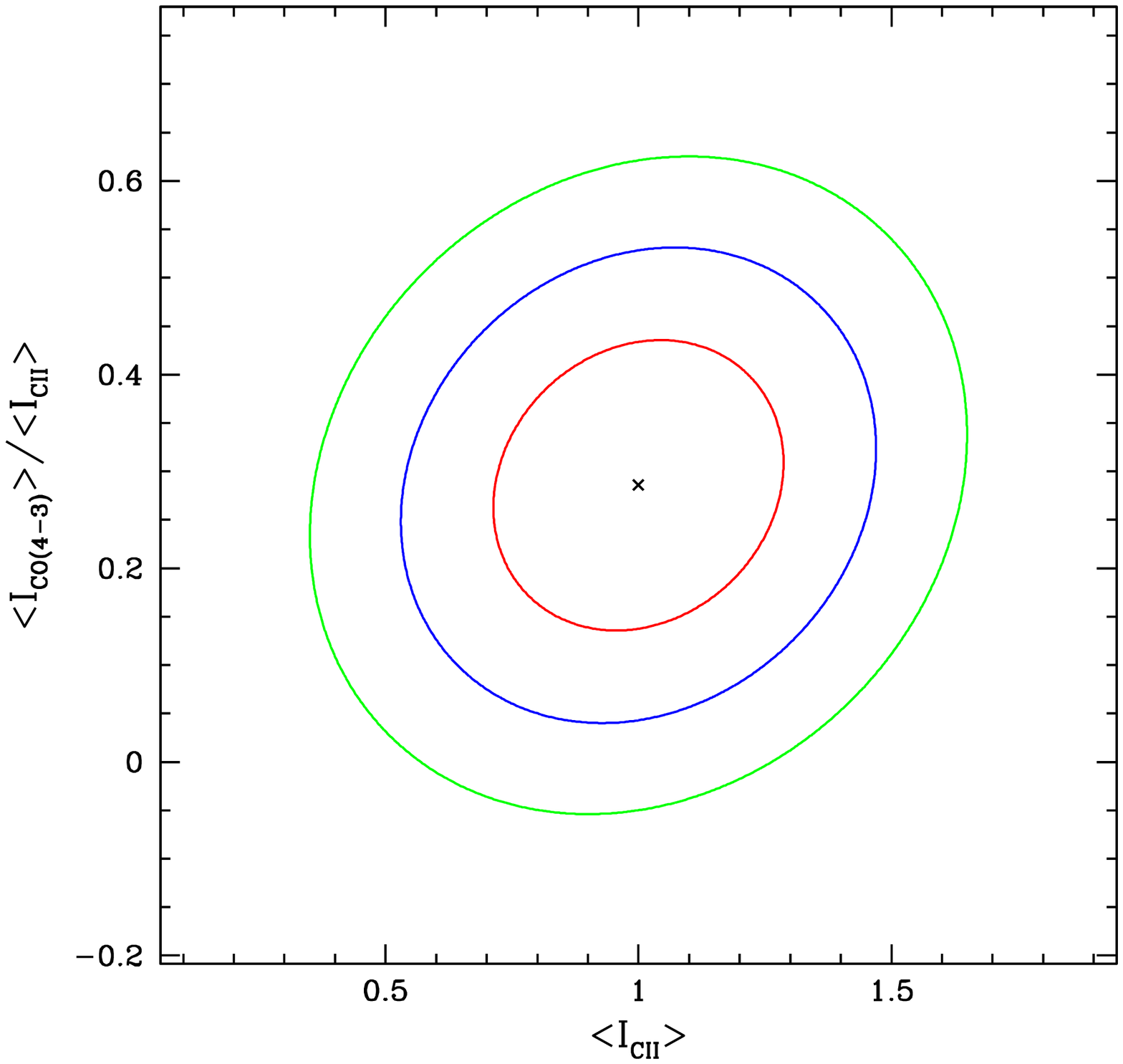}
\includegraphics[width=0.45\textwidth]{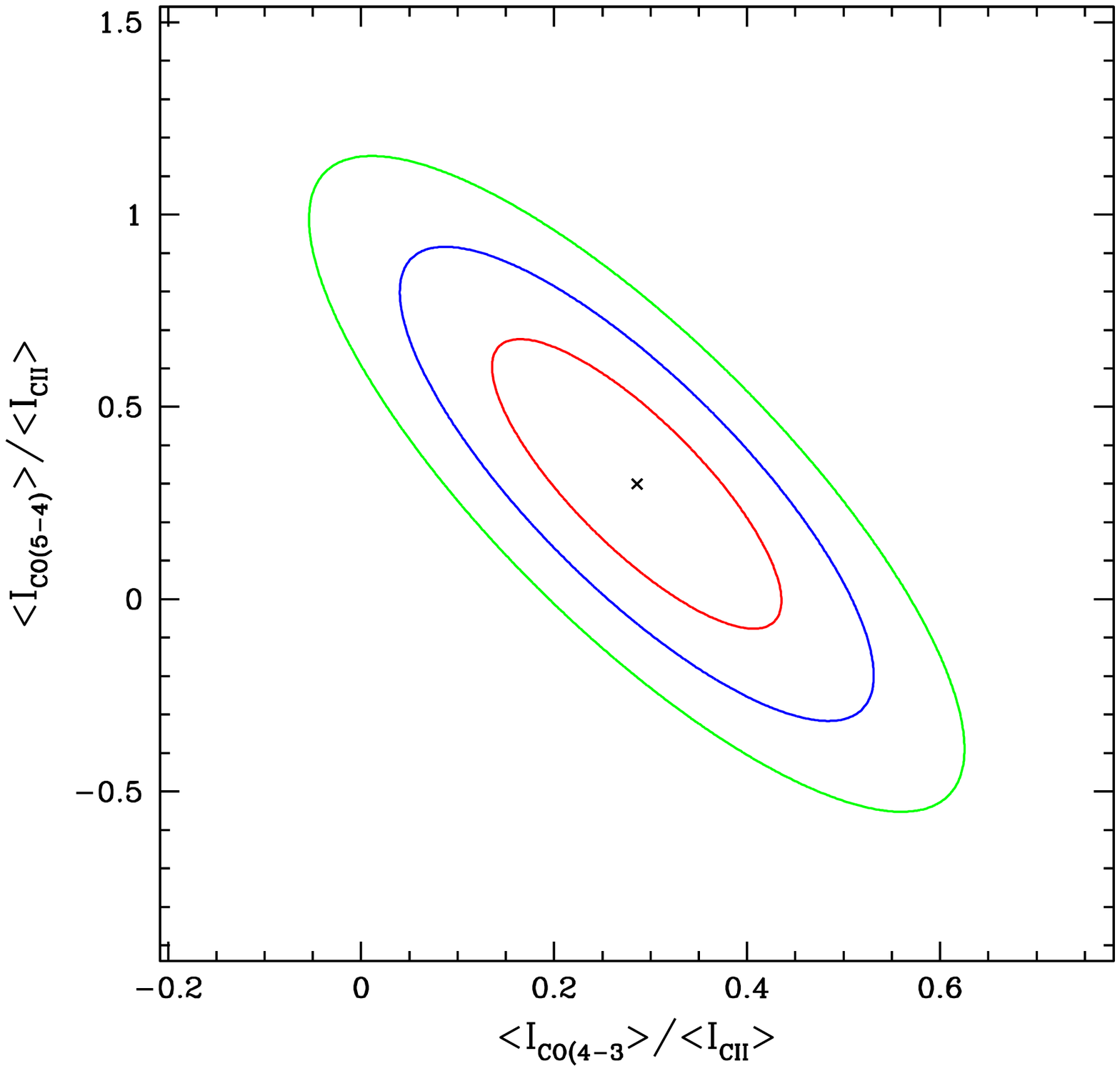}
\caption{Forecasted constraints on the average specific intensity of two of the interloper emission lines. {\em Top:} Constraints in the $\avg{I_{CII}}-\avg{I_{CO(4-3)}}$ plane. The fractional error is
larger for the interloper emission than for the target emission. This is because the average specific intensity of the target line is larger, and because the intensity of this interloper line is highly degenerate with that of other interloper lines. {\em Bottom:} Constraints in the $\avg{I_{CO(4-3}}-\avg{I_{CO(5-4)}}$ plane. The constraint ellipses show a strong degeneracy, since increasing the strength of one interloper line may be mostly compensated by reducing the strength of another line.}
\label{fig:interloper_constraints}
\end{center}
\end{figure}

It is also helpful to examine the constraints on the interloper emission parameters. Some example confidence intervals are shown in Fig.~\ref{fig:interloper_constraints}. The left hand panel shows the joint forecasted constraints in the $\avg{I_{CII}}-\avg{I_{CO(4-3)}}$  plane. This plane is of special interest because our model predicts that emission in the CO(4-3) line actually provides the largest contribution
to the total power spectrum (Fig.~\ref{fig:interlop_imp}). Interestingly, the constraints on the CO(4-3) intensity and the [CII] intensity show little degeneracy. This is actually unsurprising given the
differing angular dependence of the Fisher matrix derivatives illustrated in Fig.~\ref{fig:fisher_examp}, and the sensitive hypothetical survey we consider.
However, the different interloper lines themselves {\em are} rather degenerate with each other. This higher level of degeneracy results because the pairs of interloper lines are much closer together in redshift than the interloper-target pairs. As a result, the interloper pairs have similar distortion factors, $\alpha_\parallel$ and $\alpha_\perp$, and their power spectra hence show almost the same angular dependence. For example, the right hand panel of Fig.~\ref{fig:interloper_constraints} gives confidence intervals in the $\avg{I_{CO(4-3)}}-\avg{I_{CO(5-4)}}$
plane, and this reveals the expected strong anti-correlation between the emission in these two lines. Quantitatively, the correlation coefficient in this plane is $\rho = -0.81$. After marginalizing over all of the interloper parameters, the error bars on the average intensity of each interloper line are large: in our fiducial case, we only expect a greater than $2-3-\sigma$ detection of $\avg{I_{CO(4-3)}}$, even though
we obtain a significant detection ($ \geq 5-\sigma$) of the [CII] target emission line. For reference, our usual survey numbers forecast detections of the specific intensity in the CO(3-2) and CO(5-4) lines 
at only slightly better than $1-\sigma$, while the CO(6-5) specific intensity is still less detectable. 

In summary, the angular dependence of the emission fluctuations can be used to separate the target and interloper emission fluctuations if the noise power spectrum is sufficiently small.
Since the main goal is to extract information about the target [CII] emission, perhaps it is not a big concern that the individual CO interlopers are themselves somewhat degenerate, and the constraints
on these parameters are weaker. However, further checks seem valuable given that our approach relies on having a good model for each source of emission fluctuations.

\section{Cross-Correlating with LSS Tracers} \label{sec:xcorr_lss}
\label{sec:cross}

Fortunately, there are other approaches we can pursue as further cross checks on the analysis of the previous section, some of which should enable separate constraints on each interloper line. 
First, we can correlate the intensity mapping data
cubes with spectroscopic galaxy and/or quasar catalogues at the interloper redshift \citep{Silva:2014ira}. We expect that by the time [CII] intensity mapping experiments are underway, there will be other extensive large-scale structure surveys, spanning large fields of view and overlapping in redshift with the prominent CO interloper transitions.  We can use cross-correlations with LSS tracers at different redshifts to separately constrain the parameters of each of the various CO interloper lines. 

For instance, consider the cross power spectrum between interloper line $j$ and the abundance of spectroscopic galaxies at the same redshift, $z_j$. Suppose the average bias of these tracer galaxies is 
$\avg{b_{\rm gal}}$. In order to extract the cross spectrum of interest, it is convenient to convert from angles and wavelengths to co-moving units assuming the interloper redshift $z_j$, rather than the target [CII] redshift, $z_t$. The target line and the other interloper lines will not contribute on average to the cross spectrum with the LSS tracer at $z_j$ since these lines originate at significantly different redshifts, but they will contribute to the {\em variance} of the cross spectrum, as we will describe. For this purpose, the total power spectrum of intensity fluctuations from line emission is computed along the lines of 
Eq.~\ref{eq:power_warp_tot}, except that the warping is now relative to the coordinates of an interloper at redshift $z_j$. The cross power spectrum with the galaxy tracer field is then:
\begin{align}
P_x(k,\mu) =& \avg{I_j} \avg{b_j} \avg{b_{\rm gal}} \left(1 + \beta_j \mu^2 \right) \left(1 + \beta_{\rm gal} \mu^2 \right) \nonumber \\
& \times D\left[\mu k \sigma_p(z_j)\right] P_{\rho}(k,z_j),
\label{eq:pcross}
\end{align}
where $\beta_j = f_\Omega(z_j)/\avg{b_j}$ and $\beta_{\rm gal} = f_\Omega(z_j)/\avg{b_{\rm gal}}$ are the Kaiser parameters for the interloper line and the galaxy density field, respectively.  
We have assumed here that the finger-of-god suppression has an identical form for the IM galaxies and for the LSS tracer population at the same redshift; although this is unlikely true in detail,
we expect this simplification to have little impact on our results.
Here for simplicity we have also assumed that the CO emitting populations and the tracer galaxies are largely disparate populations; otherwise, there should be an additional shot-noise term in Eq.~\ref{eq:pcross}. 
In any case, if the cross spectrum can be measured accurately enough we can infer constraints on $\avg{I_j}$, $\avg{b_j}$, and $\avg{b_{\rm gal}}$, or at least their overall product. One final caveat here, however, is
that we have not included a stochasticity parameter ``r'' in the above equation and so we are implicitly assuming that the galaxies and interloper populations are perfect tracers of large scale structure on the scales of interest
for this measurement.
In addition, the auto spectrum of the tracer galaxies may be used to measure $\avg{b_{\rm gal}}$.  Ideally, future LSS surveys will provide tracer galaxy or quasar samples at the redshifts of each of the prominent CO interlopers. These measurements can then be combined with the angular dependence of the intensity auto spectrum, to further separate the interloper contaminants from the target emission fluctuations.

\begin{figure}
\begin{center}
\includegraphics[width=9cm]{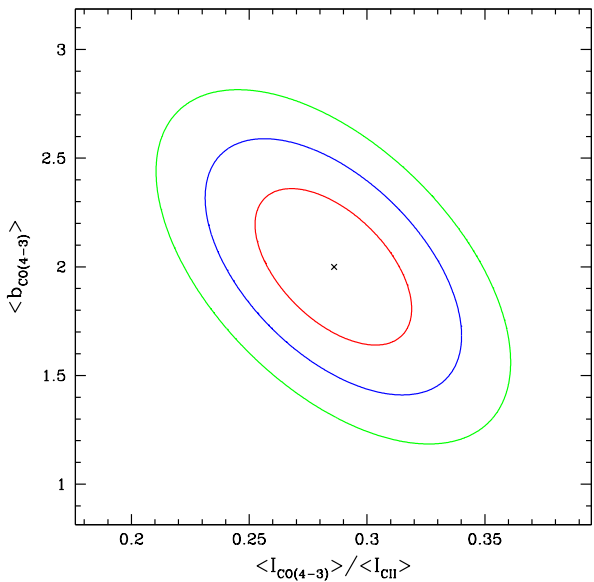}
\caption{Constraints on the parameters of an interloper line (CO(4-3)) from cross-correlating with a large-scale structure tracer at the relevant redshift ($z_j = 0.88$). A $10\%$ prior on $\avg{b_{\rm gal}}$ has been incorporated since this
parameter will be constrained from measuring the {\em auto spectrum} of the tracer galaxies. Here $\avg{I_{\rm CO(4-3)}}$ is in units of $\avg{I_{\rm CII}(z=7)}$, and our fiducial IM noise power
spectrum has been assumed.}
\label{fig:cross_interloper}
\end{center}
\end{figure}

The variance (per mode) of the cross-spectrum is given by:
\begin{align}
{\rm var}\left[P_x(k,\mu)\right] =& \Bigg[P_x^2 + \bigg(P_{\rm tot}(k,\mu) + P_N(k,\mu)\bigg) \nonumber \\ 
& \times \left(P_{\rm gal}(k,\mu) + \frac{1}{n_{\rm gal}}\right)\Bigg],
\label{eq:px_var}
\end{align}
where $P_{\rm tot}(k,\mu)$ is the total line intensity power spectrum, except computed here with the distortion factors considered relative to the coordinates at the interloper redshift $z_j$, $P_N(k,\mu)$ is
the noise power spectrum for the intensity mapping survey (Eq.~\ref{eq:pnoise}), while $P_{\rm gal}$ and $1/n_{\rm gal}$ are the clustering and shot-noise terms for the tracer galaxies.  
After specifying the properties of our tracer galaxies and the survey parameters, the cross spectrum Fisher matrix, $F^x_{ij}$, may be computed along the lines of Eq.~\ref{eq:fisher_ij}:
\begin{align}
F^x_{ij} =& \int_{\mu_{\rm min}}^{\mu_{\rm max}} d \mu \nonumber \\
& \times  \int_{k_{\rm min}}^{k_{\rm max}} \frac{dk k^2 V_s}{4 \pi^2} \frac{\partial P_x(k,\mu)}{\partial q_i} \frac{\partial P_x(k,\mu)}{\partial q_j} \frac{1}{{\rm var}[P_x(k,\mu)]}.
\label{eq:fisher_x_ij}
\end{align}
Here the parameter vector is specified by just three components: $q_\alpha = \{\avg{I_j},\avg{b_j},\avg{b_{\rm gal}} \}$.  

As an example of the cleaning that may be feasible with future data sets, we consider surveys for narrow emission-line galaxies using the Dark Energy Spectroscopic Instrument (DESI) \citep{Levi:2013gra}. We suppose that the entire volume of the intensity mapping survey is contained within the DESI narrow emission line galaxy survey, which is plausible given that DESI will cover a large-fraction of the full sky. In this case, the number of modes surveyed and the spatial and spectral resolution of the cross spectrum measurement are entirely limited by the intensity mapping survey specifications and the only additional relevant  parameters for our Fisher matrix forecasts are the tracer galaxy bias parameters (this fixes $P_{\rm gal}(k,\mu)$ in our linear biasing model), and the abundance $n_{\rm gal}$ which determines the shot-noise contribution to the variance for the DESI galaxies. We adopt the abundance of narrow emission-line galaxies that may be observed by DESI as reported in \citet{Levi:2013gra}. In this case, near the redshift of the CO(4-3) interloper emission, we expect a number density
of $n_{\rm gal} = 5.2 \times 10^{-4} h^3$ Mpc$^{-3}$.  The expected abundance of DESI tracer galaxies at the redshifts of the other prominent interlopers are comparable. 
Finally, we would like to account for the constraint that will be possible on $\avg{b_{\rm gal}}$ from a measurement of the auto-spectrum of the tracer galaxy survey. Note that the DESI emission line galaxies will themselves suffer from interloper contamination (e.g. \citealt{Pullen:2015yba}) 
and this will need to cleaned in order to measure the auto-spectrum and $b_{\rm gal}$. Rather than investigate this in detail here, for simplicity we suppose that $b_{\rm gal}$ is measured to $10\%$ 
fractional accuracy. We believe this is conservative.
This is then incorporated as a prior in the cross-spectrum Fisher matrix calculation (Eq.~\ref{eq:fisher_x_ij}). For the bias of the tracer galaxies, we adopt a central value 
of $b_{\rm gal}=2.5$. 

Fig.~\ref{fig:cross_interloper} shows an example of the constraints that may be obtained for the case of CO(4-3) interloper line emission. Evidently, the cross spectrum with the DESI narrow emission line galaxy sample should allow significantly tighter constraints on $\avg{I_{\rm CO(4-3)}}$ than from the total intensity mapping auto spectrum. For our fiducial assumptions, the $1-\sigma$ fractional error bar
on $\avg{I_{\rm CO(4-3)}}$ improves by a factor of more than four. Similar measurements should be possible for each of the other CO interloper transitions. These cross spectrum measurements should be useful both as a consistency check on the interloper modeling, and can be used in combination with the total intensity mapping auto spectrum to reduce error bars on the target emission parameters. 
Quantitatively, we can incorporate the DESI-like cross spectrum constraints on the specific intensity of the interloper lines as ($1-\sigma$) priors in our auto-spectrum Fisher matrix calculations. Doing 
this, we find
that the error bars on the [CII] specific intensity and bias shrink by a factor of $1.5$ and $1.4$ respectively. Although these numbers are indicative, the precise gain will depend on the noise power
spectrum in the IM experiment and on how accurately the auto-spectra of the DESI galaxies are measured. 

\section{Cross Spectrum with Other Lines} \label{sec:xcorr_lines}
\label{sec:cross_lines}

Finally, an additional approach to help confirm the presence of target [CII] emission fluctuations is to cross-correlate with a data cube centered on a different frequency that contains emission from another line
at the same redshift (e.g. \citealt{Visbal10}). Indeed, this measurement may potentially be done with the same data set. For example, the hypothetical [CII] survey discussed in \citet{Silva:2014ira} spans 200-300 GHz. In addition to the [CII] $158 \mu$ m
line at $z=7$, the same survey should include [OI] emission at $z=7$ with a rest frame wavelength of $146 \mu$ m, at an observed frequency of $\nu_{\rm obs}=259$ GHz. Further, just outside the fiducial range
spanned by this hypothetical survey is an [NII] $122 \mu$ m emission line at $z=7$, $\nu_{\rm obs} = 308$ GHz. The cross spectrum between the [CII] and [OI] data cubes, for example, should follow
\begin{align}
P_{x, {\rm CII-OI}}(k, \mu) =& \avg{I_t} \avg{I_{\rm OI}} \avg{b_t} \avg{b_{\rm OI}} \left(1 + \beta_t \mu^2 \right) \left(1 + \beta_{\rm Ol} \mu^2 \right) \nonumber \\
&\times  D\left[\mu k \sigma_p(z_t)\right] P_{\rho}(k,z_t) +
P_{\rm shot, CII-OI}
\label{eq:px_cii_oi}
\end{align}
where $\avg{I_{\rm OI}}$ and $\avg{b_{\rm OI}}$ denote the specific intensity and linear bias factor of the [OI] emitters that lie at the same redshift as the [CII] emission, and the other symbols have their usual meanings. Similar to Eq.~\ref{eq:pcross}, we assume that the finger-of-god suppression has an identical form for each set of emitters. In what follows, we neglect the shot-noise term, $P_{\rm shot, CII-OI}$. Strictly speaking, this is only correct in the limit that disparate populations of sources produce the [CII] and [OI] emission. However our sensitivity here is coming from large scales where the shot-noise
contribution should be small, so we don't expect neglecting it to impact our estimates. 

Here we consider using the cross-spectrum between [CII] and [OI] as a test of the high redshift origin of a potential [CII] contribution to the intensity mapping data cube. For this purpose, we define $A=\avg{I_t} \avg{I_{\rm OI}} \avg{b_t} \avg{b_{\rm OI}}$ and consider the significance at which $A$ can be shown to be non-zero. Here the relevant variance is:
\begin{align}
{\rm var}\left[P_{x,{\rm CII-OI}}(k,\mu)\right] & = 
 \Bigg[P_{x,{\rm CII-OI}}^2  \nonumber \\
& + \bigg(P_{\rm tot, CII}(k,\mu) + P_{N, {\rm CII}}(k,\mu)\bigg) \nonumber \\ 
& \times \bigg(P_{\rm tot, OI}(k,\mu) + P_{N, {\rm OI}}(k,\mu)\bigg)\Bigg],
\label{eq:px_var_cii_oi}
\end{align}
where $P_{\rm tot, OI}(k,\mu)$ is the total [OI] signal auto spectrum, including the interlopers for this line. For simplicity, we approximate the interloper power contamination to the [OI] line as identical to that of the [CII] line. This should be a good but imperfect approximation, since the two lines lie at fairly similar observed frequencies. Likewise, we approximate the noise power spectrum as identical at the observing frequencies centered around each of the [CII] and [OI] lines.
Based on the local relation between line luminosity and star-formation rate in \citet{Visbal10} and using Eq.~\ref{eq:i_avg}, 
we infer that $\avg{I_{\rm OI}}(z=7) = 0.05 \avg{I_{\rm CII}}(z=7)$. We can then estimate the total signal to noise at which the single parameter, $A$, may be
detected using Eqs.~\ref{eq:px_cii_oi} and ~\ref{eq:px_var_cii_oi}. For our fiducial numbers we find that the cross spectrum may be detected at $8.6-\sigma$ significance, and so considering the cross spectrum
between the two lines seems promising. If the frequency range can be extended somewhat, the cross spectrum between [CII] and [NII] might be detectable. In fact, based on the local line-luminosity
star formation rate correlation tabulated in \citet{Visbal10} we expect this correlation to be more detectable than that between [CII] and [OI]: using the numbers in \citet{Visbal10} gives a $17-\sigma$ 
detection forecast. However, assuming the local relation is especially suspect for [NII]: there is unlikely to be enough prior star formation to build up a significant nitrogen abundance at the high
redshifts of interest here \citep{Suginohara:1998ti}. 

Unfortunately -- for our fiducial survey numbers -- we don't expect significant detections of the auto spectra in [OI] or [NII] given the large interloper ``noise'' and the lower expected specific intensity in these lines. Consequently, a measurement
of the cross-spectrum between [CII] and [OI] and/or [NII] can help establish the high redshift origin of a possible [CII] signal, but it won't provide a full check on the values of $\avg{I_t}$, $\avg{b_t}$ inferred from
the [CII] auto spectrum, since the bias and intensity of the [OI] and/or [NII] emission will remain uncertain.

\section{Conclusions} \label{sec:conclusions}

Line confusion provides an important systematic concern for many intensity mapping surveys and for some traditional surveys targeting emission-line galaxies. Interloper line emission will likely be
especially strong in future intensity mapping surveys aimed at detecting reionization-era signals in the [CII] and Ly-$\alpha$ lines. Here we developed an approach to fit-out interloper contamination
at the power spectrum level, using the fact that the interloper contribution to the emission power spectrum will have a distinctive anisotropy that results when the target redshift is assumed in mapping from
frequency and angle to co-moving units.

We applied this to the case of a $z=7$ [CII] intensity mapping experiment, in which the $z \sim 7$ signal fluctuations are expected to be smaller than the combined emission fluctuations from several
CO interloper lines. In the limit of low noise power, the interloper fluctuations can be separated from the [CII] power spectrum signal. A more sensitive instrument than currently planned is however  required. 
In the near term, it would be interesting to investigate whether the power spectrum anisotropy technique advocated here may be fruitfully combined with a masking approach. Additional careful work
is required to study this; in this context, it is crucial to examine optical and infrared tracers to quantify whether they may serve as faithful proxies for the CO interloper emission.
We therefore defer this to future work.

We also explored how the intensity mapping data cube may be cross-correlated with large scale structure tracers to extract the properties of likely interloper lines. We showed that emission-line
galaxy samples from DESI will be a good data set for cross-correlations, allowing one to extract CO interloper 
properties for $z \sim 7$ [CII] emission surveys. Finally, we briefly considered the cross-correlation between two
different fine structure lines at the same redshift; this can help verify the high redshift origin if a possible signal is seen in the $z \sim 7$ [CII] auto spectrum. For all of these studies, it will be important
to further consider foreground contamination systematics. Specifically, additional work is needed to quantify the impact of mode-mixing on efforts to measure the angular dependence of the [CII] power
spectrum. It will also be important to quantify how correlated the foregrounds for different tracer lines -- such as [CII] and [OI] -- are.

In any case, intensity mapping is a potentially powerful approach for tracing large-scale structure at early times and may capture the collective impact of  sources that are undetectable using traditional
means. Although interloper contamination is a concern for many of these measurements, it may be circumvented using a combination of techniques, including the power spectrum anisotropy approach
considered here.

\section*{Acknowledgements}

AL and JT were supported in part by NASA grant NNX12AC97G. We thank James Aguirre for helpful conversations and Yun-Ting Cheng for useful discussions and comments on a draft manuscript.

\bibliography{references}

\end{document}